\documentclass[letterpaper,twocolumn,english,prl,groupedaddress]{revtex4}
\usepackage[T1]{fontenc}
\usepackage[latin9]{inputenc}
\usepackage{color}
\usepackage{verbatim}
\usepackage{textcomp}
\usepackage{mathrsfs}
\usepackage{mathtools}
\usepackage{amsmath}
\usepackage{amssymb}
\usepackage{graphicx}

\makeatletter

\pdfpageheight\paperheight
\pdfpagewidth\paperwidth

\providecommand{\tabularnewline}{\\}

\@ifundefined{textcolor}{}
{
 \definecolor{BLACK}{gray}{0}
 \definecolor{WHITE}{gray}{1}
 \definecolor{RED}{rgb}{1,0,0}
 \definecolor{GREEN}{rgb}{0,1,0}
 \definecolor{BLUE}{rgb}{0,0,1}
 \definecolor{CYAN}{cmyk}{1,0,0,0}
 \definecolor{MAGENTA}{cmyk}{0,1,0,0}
 \definecolor{YELLOW}{cmyk}{0,0,1,0}
}

\RequirePackage[colorlinks=true,linkcolor=blue]{hyperref}
\usepackage{dsfont}

\AtBeginDocument{
  
}

\makeatother

\usepackage{babel}

\begin{document}

\title{Acoustic Traps and Lattices for Electrons in Semiconductors}

\author{M. J. A. Schuetz,$^{1,2,*}$ J. Knörzer,$^{1,*}$ G. Giedke,$^{3,4}$
L. M. K. Vandersypen,$^{5}$ M. D. Lukin,$^{2}$ and J. I. Cirac$^{1}$ }

\affiliation{$^{1}$Max-Planck-Institut für Quantenoptik, Hans-Kopfermann-Str.
1, 85748 Garching, Germany}
\affiliation{$^{2}$Physics Department, Harvard University, Cambridge, MA 02318,
USA}
\affiliation{$^{3}$Donostia International Physics Center, Paseo Manuel de Lardizabal
4, E-20018 San Sebastián, Spain}

\affiliation{$^{4}$Ikerbasque Foundation for Science, Maria Diaz de Haro 3, E-48013
Bilbao, Spain}
\affiliation{$^{5}$Kavli Institute of NanoScience, TU Delft, P.O. Box 5046, 2600
GA Delft, The Netherlands}

\thanks{These authors have contributed equally to this work.}

\date{\today}

\begin{abstract}
We propose and analyze a solid-state platform based on surface
acoustic waves (SAWs) for trapping, cooling and controlling (charged)
particles, as well as the simulation of quantum many-body systems.  
We develop a general theoretical framework demonstrating the emergence
of effective time-\textit{independent} acoustic trapping potentials for
particles in two- or one-dimensional structures. 
As our main example we discuss in detail the generation and applications
of  a stationary, but movable acoustic pseudo-lattice (AL)  with lattice
parameters that are reconfigurable \textit{in situ}. 
We identify the relevant figures of merit, discuss potential
experimental platforms for a faithful implementation of such an acoustic
lattice, and provide estimates for typical system parameters.
With a projected lattice spacing on the scale of $\sim100\mathrm{nm}$,
this approach allows for relatively large energy scales in the
realization of fermionic Hubbard models, with the ultimate prospect of
entering the low temperature, strong interaction regime.  
Experimental imperfections as well as read-out schemes are discussed. 
\end{abstract}

\maketitle

\section{Introduction }

The ability to trap and control particles with the help of
well-controlled electromagnetic fields has led to revolutionary advances
in the fields of biology, condensed-matter physics, high-precision
spectroscopy and quantum information, enabling unprecedented control
both in the study of isolated single particles as well as few- and many-body
systems subject to controlled and tunable interactions. 
Prominent examples range from using optical tweezers for probing the
mechanical properties of DNA \cite{ashkin00, lang03}, to the
realizations of Bose-Einstein condensates
\cite{anderson95, bradley95, davis95} and numerous breakthrough
investigations of strongly-correlated quantum many-body  systems with
both trapped ions \cite{blatt12} and ultracold atoms in optical lattices
\cite{bloch08, bloch12}. 
At the same time, the ever improving control of materials and
fabrication of semiconductor nanostructures has led to a proliferation
of quasi-particles in such systems  and a quest to trap and isolate them 
in order to gain deeper insights into their properties and interactions. 
While quantum dots have been developed into excellent traps for charged
and neutral quasiparticles and have contributed to a wealth of exciting
insights \cite{hanson07}, scaling them to the many-body regime remains
either a fabrication or operational challenge. 
This motivates our
search for trapping mechanisms that bring the generality and flexibility
of optical lattices to the solid-state setting.

While an optical approach may be feasible \cite{schuetz10},
surface-acoustic waves (SAWs) have recently been used in a range of
exciting experiments to trap electrons
\cite{stotz05,hermelin11,mcNeill11,cunningham99,ford17} or excitons \cite{lima05} in
\emph{moving} potentials. 
When following this approach, however, particles are typically lost on a
relatively fast timescale of $\sim10$ns, as a consequence of finite
sample sizes and propagation speeds set by the speed of sound to
$\sim 3\times 10^{3}\mathrm{m/s}$.
Inspired by these experiments, here we propose and analyze engineered
stationary and quasi-stationary (movable) \textit{acoustic} trapping
potentials and acoustic lattices (ALs) as a generic strategy for
trapping, cooling and controlling quasi-particles as well as a potential
on-chip, solid-state platform for the simulation of quantum many-body
systems.
While in this work we use the generation of an effective standing-wave
lattice for electrons as the main example of our technique, our
theoretical approach generalizes immediately to other trap
configurations. 
In particular, focused SAWs \cite{lima03} might allow for the generation 
of quasi zero-dimensional traps for electrons akin to optical tweezers, 
thereby entering a new parameter regime in the context of \textit{acoustic tweezers}; 
so far, the latter have been used only in a high-temperature, classical regime to trap and
manipulate microparticles immersed in fluids above the SAW-carrying solid \cite{ding13}.

\begin{figure}[b]
\includegraphics[width=0.9\columnwidth]{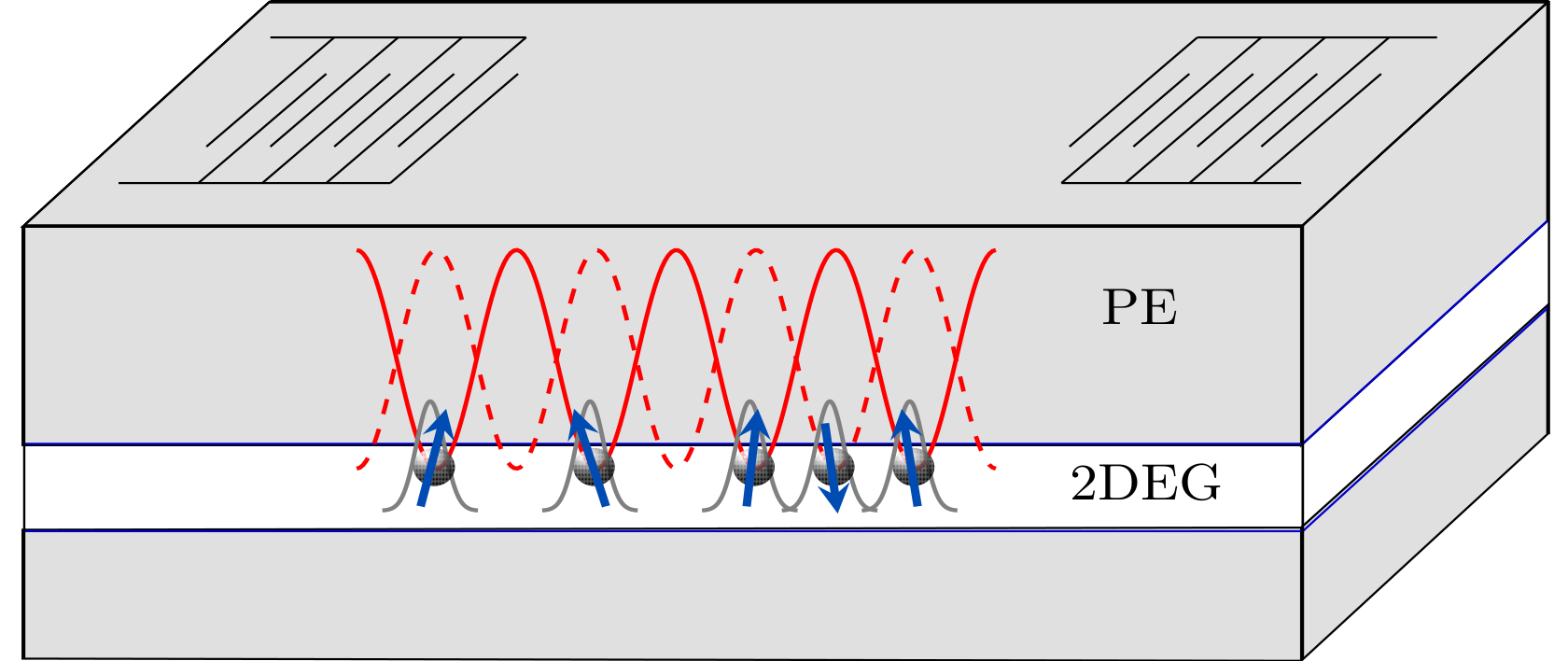}
\caption{\label{fig:setup}(color online).
Exemplary schematic illustration of the setup. In a piezo-electric solid
(PE) counter-propagating SAWs (as induced by standard IDTs deposited on
the surface \cite{datta86,morgan07}) generate a \textit{time-dependent},
periodic electric potential for electrons confined in a conventional
two-dimensional electron gas (2DEG). 
If the SAW frequency $\omega/2\pi=v_{s}/\lambda$ is sufficiently high
(as specified in the main text), the electron's potential landscape can
effectively be described by a \textit{time-independent} pseudo-lattice
with a lattice spacing $a=\lambda/2$.
The potential depth (lattice spacing) can be controlled conveniently via
the power (frequency) applied to the IDTs; while an additional screening
layer (not shown) allows for tuning the strength of the Coulomb
interaction between the particles \cite{byrnes07}.
In more complex structures, the setup can consist of multiple layers on
top of some substrate.}
\end{figure}

Our basic scheme involves counter-propagating SAWs that are launched in
opposite directions from two (or more) standard interdigital transducers
(IDTs) \cite{datta86,morgan07} patterned either directly onto a
piezoelectric substrate such as GaAs or  on some piezoelectric island as
demonstrated for example in Ref.~\cite{cerda-mendez10}; for a schematic
illustration compare Fig.\ref{fig:setup}. 
Because of the intrinsic piezoelectric property of the material, the
SAWs are accompanied by a (time-dependent) periodic electric potential
and strain field, generating a well-controlled potential landscape (of
the same spatial and temporal periodicity) for electrons confined in
conventional quantum wells or purely two-dimensional crystals such as 
transition metal dichalcogenides (TMDCs), with a periodicity on the
order of $\sim100\mathrm{nm}$ for SAW frequencies of
$\sim20\mathrm{GHz}$ \cite{kukushkin04}.
Based on a perturbative Floquet approach, we show that the electron's
potential landscape can effectively be described by a
\textit{time-independent} pseudo-lattice with a lattice spacing
$\sim a=\lambda/2$, provided that certain conditions are fulfilled (as
specified below).
Intuitively, the occurrence of such an effective
time-independent potential can be understood from the fact that
sufficiently heavy electrons cannot adiabatically follow a rapidly
oscillating force as created by the SAW-induced electric potential;
therefore,   the electron will effectively be trapped close to the
potential minimum  if its wavefunction spreads slowly enough such that
it is still close to its original position after one oscillation period
of the SAW field.
We identify the relevant figures of merit for this novel setup [cf. Eq.(\ref{eq:energy-scale-Es})]
and show how the system parameters can be engineered and dynamically tuned.
As a guideline for an experimental realization
of the proposed setup, we derive a set of self-consistency requirements
which allows us to make clear predictions about the material properties
needed for a faithful implementation. Consequently, we identify strategies
to meet these requirements with state-of-the-art experimental techniques
and suitable material choices. Concerning the latter, we
analyze the viability of different heterostructures with high effective
electron masses which support high-velocity sound waves, e.g. AlN/diamond
or, alternatively, TMDCs
such as $\mathrm{MoS}_{2}$ or $\mathrm{WSe}_{2}$.
While we discuss the relevant decoherence mechanisms as well as other
relevant experimental imperfections for specific systems, the very basic
principles of our approach should be of broad applicability to various
physical solid-state platforms. 
In particular, thanks to the generic nature of our analysis
and the variety of fields (strain, electric, magnetic) that potentially
accompany SAWs, our framework is readily applicable to a broad class of
(quasi-)particles, including for example electrons, holes, trions and
excitons.
While our theoretical treatment is (to some extent) reminiscent of trapped ions, 
allowing us to capitalize on ideas and results from this well-developed field of research, 
we show that the emergent effective dynamics can be captured by the Fermi-Hubbard model, 
very much like for fermionic ultra-cold atoms in optical lattices, 
albeit in unprecedented parameter regimes, 
because of ultra-high charge-to-mass ratios and naturally long-ranged Coulomb interactions.
Our approach provides an alternative to standard (gate-defined) quantum
dots, providing a highly regular periodicity simply set by the SAW
wavelength,  with minimal fabrication requirements (without any further
gate patterning),  and the potential to deterministically move around
the acoustically-defined quantum dots by simply changing the phase of
the excitation applied to the IDTs.
Also, our trapped-ion-inspired pseudo-potential approach makes our
proposal significantly different from previous theoretical
\cite{byrnes07} and experimental investigations
\cite{cerda-mendez10, lima05}, where particles trapped inside a dynamic,
\textit{moving} AL (rather than a quasi-stationary, \textit{standing}
AL, as considered here) are inevitably lost within a rather short
timescale $\sim10\mathrm{ns}$.

\section{Theoretical Framework \label{sec:theory}}

In this section we first develop a general theoretical framework
describing particles in low-dimensional semiconductor structures in the
presence of (SAW-induced) high-frequency standing waves. 
We employ both classical and quantum-mechanical tools in order to
identify the relevant figures of merit and specify the conditions for
the validity of our theoretical framework.  
The experimental feasibility of our scheme will be discussed for
specific setups thereafter in Section \ref{sec:implementation}.

\textit{Surface acoustic waves}.\textemdash SAWs are phonon excitations
which propagate elastically on the surface of a solid within a depth of
roughly one wavelength $\lambda$ \cite{datta86,morgan07}.
In the case of a piezoelectric material, SAWs can be generated
electrically based on standard interdigital transducers (IDTs) deposited
on the surface, with a SAW amplitude proportional to the amplitude
(square root of the power) applied to the IDTs
\cite{datta86,morgan07,hermelin11}.
Typically, such an IDT consists of two thin-film electrodes on a
piezoelectric material, each formed by interdigitated fingers.
Whenever a radio frequency (RF) signal is applied to such an IDT, a SAW
is generated if the resonance condition $p=v_{s,\alpha}/f$ is met; here,
$p$, $v_{s,\alpha}$ and $f=\omega/2\pi$ refer to the IDT period, the
sound velocity of a particular SAW mode $\alpha$ and the applied
frequency, respectively \cite{datta86,morgan07,buy12}.
As evidenced by numerous experimental studies
\cite{hutson62,bierbaum72,wixforth86,wixforth89}, SAWs can interact with
a two-dimensional electron gas (2DEG) via the electric (and/or strain)
field accompanying this elastic wave.

\begin{figure}
\includegraphics[width=\columnwidth]{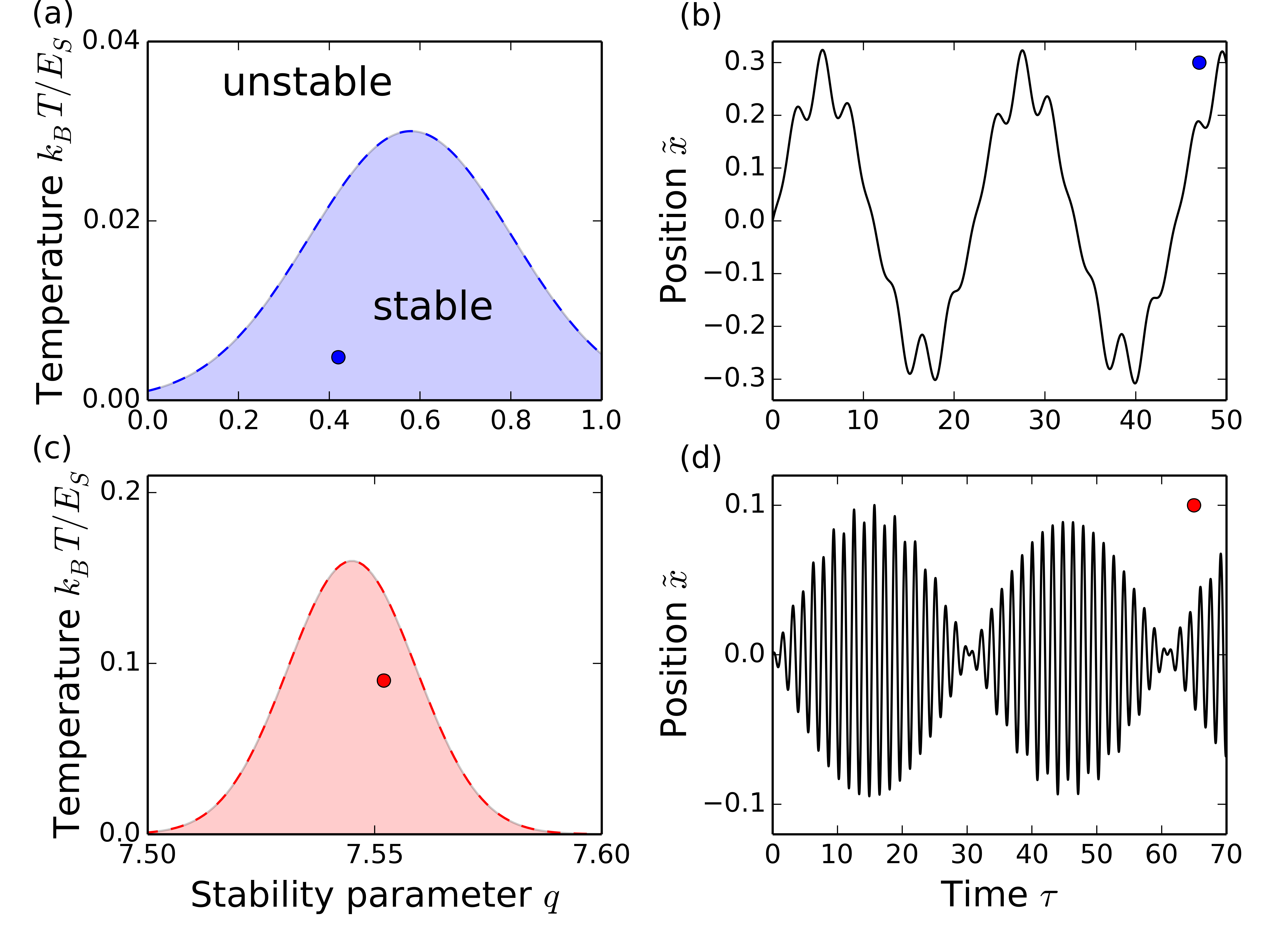}
\caption{\label{fig:stability}(color online). Approximate stability diagrams
of the classical equation of motion in the low-$q$ (upper plot) and
high-$q$ (lower plot) regimes, respectively. The dots denote trajectories
corresponding to some exemplarily chosen parameter sets $(q,\ k_{B}T/E_{S})$.}
\end{figure}

\textit{Classical analysis}.\textemdash
To illustrate our approach, let us first consider the classical dynamics
of a single, charged particle of mass $m$ (also referred to as
\textit{electron} in the following) exposed to a SAW-induced
monochromatic piezo-electric standing wave of the form
$\phi\left(x,t\right)=\phi_{0}\cos\left(kx\right)\cos\left(\omega t\right)$.
Here, $\omega=v_{s}k$ refers to the dispersion relation of a specific
SAW mode and the time-dependent potential experienced by the electron is
$V\left(x,t\right)=e\phi\left(x,t\right)$ with an amplitude
$V_{\mathrm{SAW}}=e\phi_{0}$ (where $e$ denotes the electron's charge). 
In the absence of a piezoelectric potential, a similar periodic
potential derives from the (strain-induced) deformation potential
associated with a SAW \cite{lima05}; our theoretical analysis applies to
both scenarios, as it is independent of the microscopic origin of the
SAW-induced potential
$V(x,t)=V_{\mathrm{SAW}}\cos\left(kx\right)\cos\left(\omega t\right)$. 
While the motion in the $z$-direction is frozen out for experimentally
relevant temperatures, a potential pattern of the same periodic form
could be produced in the $y$-direction using appropriately aligned pairs
of IDT's launching counter-propagating SAWs \cite{byrnes07}.
In this scenario the electron's motional degrees of freedom are
separable into two one-dimensional problems of the same structure.
Alternatively, using for example etching techniques or gate-defined
structures as described in Refs.\cite{hermelin11,mcNeill11},
effectively one-dimensional wires with strong transverse confinement in
the $y$-direction may be considered. Therefore, in any case only the
motion in the $x$-direction will be discussed in the following.
Then, in dimensionless units, where $\tilde{x}=kx$ and
$\tau=\omega t/2$, Newton's equation of motion for the electron's
position $x(t)$ reads
\begin{equation}
\frac{d^{2}\tilde{x}}{d\tau^{2}}+2q\sin(\tilde{x})\cos(2\tau)=0,\label{eq:newton}
\end{equation}
where we have introduced the (dimensionless) stability parameter
$q=V_{\text{SAW}}/E_{S}$, with the emerging energy scale 
\begin{equation}
E_{S}=mv_{s}^{2}/2,\label{eq:energy-scale-Es}
\end{equation}
that is, the classical kinetic energy of a particle with mass $m$
and velocity equal to the speed of sound $v_{s}$ of the driven SAW-mode;
as will be shown below, the energy scale $E_{S}$ turns out to be
a key figure of merit in our setup. In the Lamb-Dicke limit
$\tilde{x}\ll1$, Eq.$\ $(\ref{eq:newton}) reduces to the so-called
Mathieu equation {[}cf. Eq.\eqref{eq:mathieu}{]}, which is known to
govern the dynamics of ions in Paul traps \cite{paul90, leibfried03}.
We assess the stability of the electron's motion against thermal noise
by numerically solving Eq.$\ $(\ref{eq:newton}), for initial conditions
set as
$\tilde{x}_{0}=0,\ \tilde{v}_{0}:=[d\tilde{x}/d\tau]_{\tau=0}=\sqrt{2k_{B}T/E_{S}}$;
here, according to $mv_{0}^{2}/2=k_{B}T/2$, the initial velocity $v_{0}$
is identified with the temperature $T$ by simple equipartition.
Solutions to this problem are deemed \textit{stable} if the maximal
excursion $x_{\text{max}}$ is smaller than one half of the lattice
spacing ($\tilde x_{\text{max}} < \pi$), even for very long timescales,
and \textit{unstable} otherwise.
The results of this classification procedure are shown in
Fig.\ref{fig:stability}:
Stable (bounded) solutions can only be found for sufficiently low
temperatures (with $k_{B}T\ll E_{S}$) and certain values of the
stability parameter $q$.
In particular, in the regime $q^{2}\ll1$, $k_{B}T\ll E_{S}$ stable
trajectories $\tilde{x}(\tau)$ consist of slow harmonic oscillations at
the secular frequency $\omega_{0}/\omega\approx q/\sqrt{8}$,
superimposed with fast, small-amplitude oscillations at the driving
frequency $\omega$ (also referred to as micromotion \cite{leibfried03});
compare Fig.\ref{fig:stability}(b).
When neglecting the micromotion within the so-called pseudo-potential
approximation (as routinely done in the field of trapped ions
\cite{leibfried03}), the electron's (secular) dynamics is effectively
described by that of a time-independent harmonic oscillator with (slow)
frequency $\omega_{0}\ll\omega$; for further analytical and numerical
details we refer to Appendix \ref{app:classical}.

\textit{Quantum-mechanical Floquet analysis}.\textemdash
The results described above can be corroborated within a fully
quantum-mechanical model.
Here, the electron's dynamics are governed by the
\textit{time-dependent} Hamiltonian
\begin{equation}
H_{S}\left(t\right)=\frac{\hat{p}^{2}}{2m}+
V_{\mathrm{SAW}}\cos\left(\omega t \right)\cos\left(k\hat{x}\right),\label{eq:hamilt_sys}
\end{equation}
where $\hat{x}$ and $\hat{p}$ refer to the particle's position and
momentum operators, respectively. 
The Hamiltonian $H_{S}\left(t\right)$ satisfies $H_{S}(t+T)=H_{S}(t)$
due to the time-periodic nature of the external driving, with
$T=2\pi/\omega$. 
In a high-frequency field, where the period of the force $T$ is small
compared to all other relevant timescales, the particle's dynamics can
be approximately described by a \textit{time-independent}
Hamiltonian $H_{\mathrm{eff}}$. 
As detailed in Appendix  \ref{app:qm-floquet}, $H_{\mathrm{eff}}$ can be
calculated in a systematic expansion in the inverse of the driving
frequency $\omega$ \cite{rahav03,rahav03b}.
Then, up to second order in $\sim \omega^{-1}$, we obtain
\begin{equation}
H_{\text{eff}}=\frac{\hat{p}^{2}}{2m}+V_{0}\sin^{2}(k\hat{x}),\label{eq:hamilt_eff}
\end{equation}
where $V_{0}=\varepsilon^{2}E_{S}$, with the small parameter
$\varepsilon=q/\sqrt{8}$.
The second term $V_{\mathrm{eff}}(\hat{x})=V_{0}\sin^2(k\hat{x})$
demonstrates  the formation of an effectively \textit{time-independent},
spatially periodic acoustic lattice, with a  lattice spacing
$a=\lambda/2=\pi/k$ and potential depth $V_{0}=\varepsilon^{2}E_{S}$. 
Similar to the case for trapped ions, lattice sites are found at the
nodes of the time-dependent force
$\mathbf{F}(x,t) \sim \sin(kx)\cos(\omega t) \mathbf{x}$ associated with
the potential $V(x,t)$.
This force changes its sign on a timescale $\sim \omega^{-1}$; if this
is fast compared to the particle's dynamics $\sim \omega_{0}^{-1}$,
the particle will be dynamically trapped, because it does not have
sufficient time to react to the periodic force before this force changes
its sign again.
Within the usual harmonic approximation, where
$V_{\mathrm{eff}}(\hat{x}) \approx (m/2)\omega^{2}_{0}\hat{x}^2$, the
effective trapping frequency $\omega_{0}$ can be estimated as 
$\omega_{0}/\omega \approx q/\sqrt{8}$, which coincides exactly with the
(classical) result for the slow secular frequency $\omega_{0}$ in the
pseudopotential regime (with $q^2 \ll 1$).  
Accordingly, the AL can be rewritten as 
$V_{\mathrm{eff}}(\hat{x})=(\omega_{0}/\omega)^2E_{S}\sin^2(k\hat{x})$, 
with the first (perturbative) factor accounting for the inherent
separation of timescales between the fast driving frequency $\omega$ and
the slow secular frequency $\omega_{0}$. 
Written in this form, the effective acoustic potential
$V_{\mathrm{eff}}(\hat{x})$ is reminiscent of standard dipole traps for
ultra-cold atoms. 
Here, the effective optical potential for a two-level system driven by a 
Rabi-frequency $\Omega$ with detuning $\Delta$ in a electromagnetic
standing wave takes on the form
$V_{\mathrm{opt}}(\hat{x})=(\Omega^2/4\Delta^2)\Delta \sin^2(k\hat{x})$,
with the self-consistent requirement $\Delta \gg \Omega$. 
Therefore, 
with the pre-factor $\sim\Omega^2/4\Delta^2$ being small for self-consistency,
we can associate the role $E_{S}$ plays in the acoustical
case with the role the detuning $\Delta$ plays in the optical setting. 
Along these lines, for robust trapping it is favourable to increase the material-specific
quantity $E_{S}$, thereby achieving a larger trap depth $V_{0}$ while
keeping both the stability parameter $q=V_{\mathrm{SAW}}/E_{S}$ and thus also the
perturbative parameter $\varepsilon$ constant.
This can be well understood intuitively, since trapping due to a rapidly oscillating (SAW) field only becomes 
possible if the particle is too inert to adiabatically follow the periodically applied force: 
an electron does not significantly move away from a potential minimum if during one oscillation period of the 
SAW field its wavefunction spreads slowly enough such that it is still close to its original position when the minimum reforms. 
This simplified (pseudo-potential) picture is valid for relatively heavy electrons with high mass $m$ and sufficiently 
high driving frequency (that is, high speed of sound $v_{s}$), as captured by an elevated
sound energy $E_{S}=(m/2)v^2_{s}$.


\begin{figure}
\includegraphics[width=1\columnwidth]{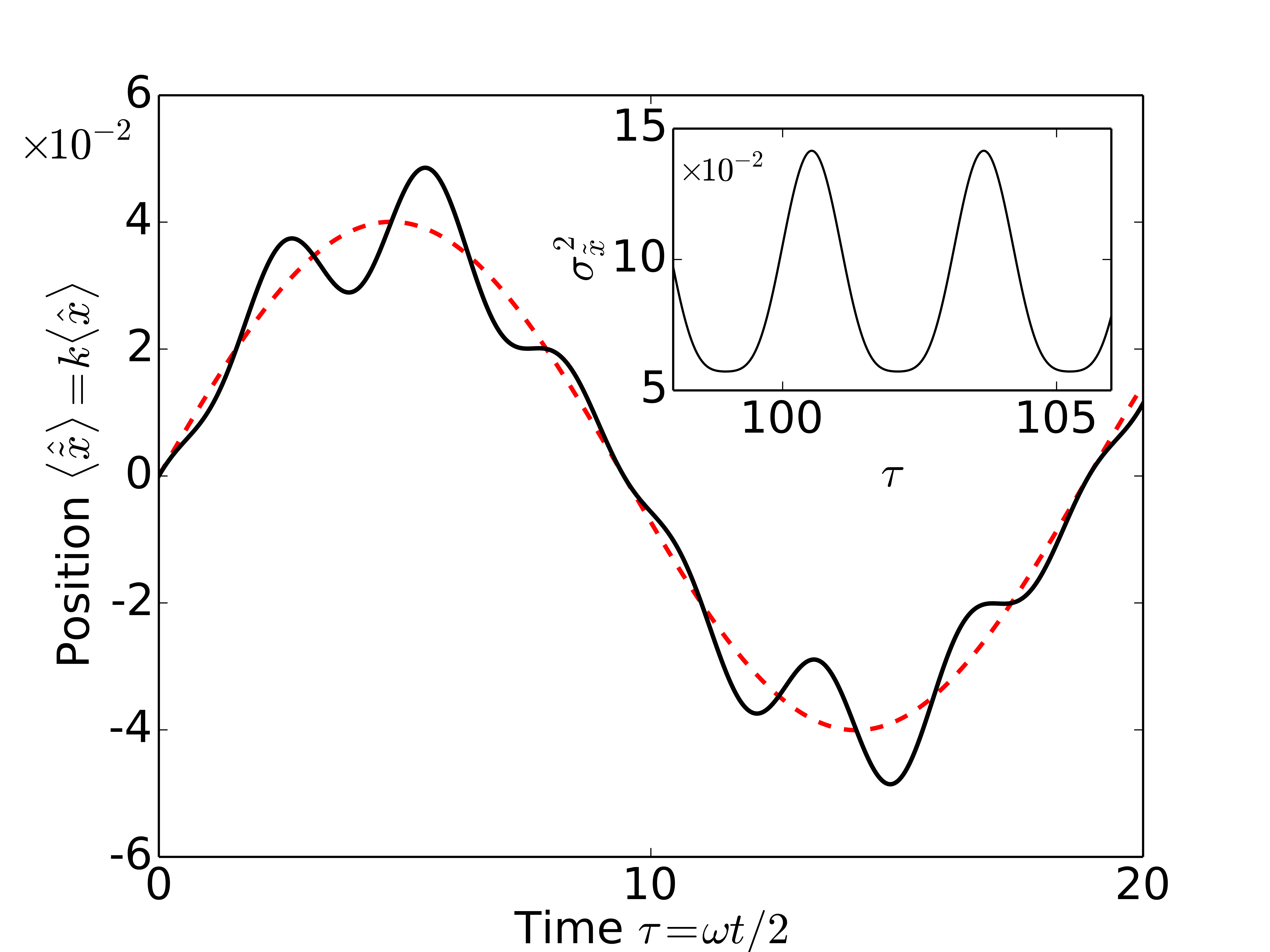}
\caption{\label{fig:simulation-gaussian-dynamics}(color online).
Exact numerical simulation
[based on Eqs.(\ref{eq:Kohler-QME-Schroedinger-picture}) and
(\ref{eq:qme-ho-maintext})] for the electron's trajectory
$\left\langle \hat{x}\right\rangle _{t}$ (solid black line), showing a
slow secular motion with frequency $\omega_{0}$ that is superimposed by
fast, small-amplitude micromotion oscillations.
When disregarding micromotion, the dynamics can approximately be
described by a simple damped harmonic oscillator with secular frequency
$\omega_{0}$ (\textcolor{black}{dashed} red line).
The initial state has been set as a coherent state with
$\langle \hat{\tilde x} \rangle = 0, \langle \hat{\tilde p} \rangle = 0.01$.
Other numerical parameters: $q=0.47$, $\gamma/\omega_{0}=10^{-3}$,
$k_{B}T/\hbar\omega_{0}=10^{-1}$, $\omega_0 / \omega \approx 0.17$.
Inset: Position variance
$\sigma_{\tilde x}^2 = \langle \hat{\tilde x}^2 \rangle - \langle \hat{\tilde x} \rangle^2$
at times when transient effects have decayed.}
\end{figure}

\textit{Cooling in the presence of micromotion}.\textemdash
While our previous discussion has exclusively focused on the
time-dependent system's dynamics, in the following we extend our studies
and introduce a dissipative model, which describes the electron's
motional coupling to the (thermal) phonon reservoir. 
For details of the derivation, we refer to Appendix
\ref{app:cooling-micromotion}.
Within one unified Born-Markov and Floquet framework, we derive an
effective quantum-master equation (QME) for the electronic motion in the
vicinity of one lattice site, fully taking into account the explicit
time-dependence of the system Hamiltonian \eqref{eq:hamilt_sys}. 
Since the quantum-state evolution due to this QME is Gaussian, one can
readily derive a closed set of equations for the first- and second-order
moments of the position and momentum observables; formally, it takes on
the form  
$\mathbf{\dot{v}}=\mathcal{M}\left(t\right)\mathbf{v}+\mathbf{C}\left(t\right)$
with
$\mathbf{v}=\left(
\left\langle \hat{x}\right\rangle _{t},
\left\langle \hat{p}\right\rangle _{t},
\left\langle \hat{x}^{2}\right\rangle _{t},
\left\langle \hat{p}^{2}\right\rangle _{t},
\left\langle \hat{x}\hat{p}+\hat{p}\hat{x}\right\rangle _{t}
\right)^{\top}$.
This equation of motion can be readily solved by numerical integration; 
a prototypical result of this procedure is displayed in
Fig.\ref{fig:simulation-gaussian-dynamics}.
In the regime $q^{2}\ll1$, our numerical findings show that
(i) the electronic motion can be described very well by a simple damped
harmonic oscillator with secular frequency $\omega_{0}$,
(ii) the electronic motion is cooled by the phonon reservoir and
(iii) the Lamb-Dicke approximation is well-satisfied.
Let us elaborate on these statements in some more detail:
(i) As evidenced by the dashed red line in
Fig.\ref{fig:simulation-gaussian-dynamics}, we find that the effective,
time-independent master equation
\begin{eqnarray}
\dot{\rho} & = & -i\omega_{0}\left[a^{\dagger}a,\rho\right]+
\gamma\left(\bar{n}_{\mathrm{th}}\left(\omega_{0}\right)+1\right)\mathscr{D}\left[a\right]\rho\nonumber \\
 &  & +\gamma\bar{n}_{\mathrm{th}}\left(\omega_{0}\right)\mathscr{D}\left[a^{\dagger}\right]\rho,
\label{eq:qme-ho-maintext}
\end{eqnarray}
captures very well the most pertinent features of the electronic
dynamics (for $q^{2}\ll1$).
Here, $\gamma$ is the effective, incoherent damping rate due to coupling
to the thermal phonon reservoir,
$\bar{n}_{\mathrm{th}}\left(\omega_{0}\right)=1/\left(\exp[\hbar \omega_{0}/k_{B}T]-1\right)$
gives the thermal occupation number of the phonon bath at frequency
$\omega_{0}$,
$\mathscr{D}\left[a\right]\rho=a\rho a^{\dagger}-(1/2)\left\{a^{\dagger}a,\rho\right\}$
denotes the standard dissipator of Lindblad form, and $a^{(\dagger)}$
refers to the usual annihilation (creation) operators for the canonical
harmonic oscillator.
As a consequence of the presence of the dissipator, the first-order
moments $\langle \hat x \rangle_t$, $\langle \hat p \rangle_t$ decay
towards zero in the asymptotic limit $t \rightarrow \infty$.
However, the second-order moments retain the periodicity of the external
driving for arbitarily long times 
(with a periodicity $\bar T = \omega T / 2 = \pi$), 
which is the signature of an emerging
quasi-stationary state
[cf. Appendix \ref{app:cooling-micromotion} for details]
and the persisting micromotion which manifests itself in the fast
oscillating dynamics of the position and momentum variances, as depicted
in the inset of Fig.\ref{fig:simulation-gaussian-dynamics}. 
(ii) As suggested by our analytical results
[cf. Appendix \ref{app:cooling-micromotion} for details], 
the phonon reservoir provides an efficient cooling mechanism for the
electron provided that the host temperature is sufficiently low, that is
$k_{B}T\ll\hbar\omega_{0}$.
The influence of the electronic micromotion on this cooling mechanism
can be condensed in the following statement: in the pseudopotential
regime (for which $q^{2}\ll1$), the expectation value for the averaged
quantum kinetic energy (over one micromotion period) features a surplus
of energy, in addition to the zero-point kinetic energy in the ground
state of $\hbar\omega_{0}/4$. 
This excess energy $\Delta_{\text{heat}}\gtrsim\hbar\omega_{0}/4$ may be
viewed as micromotion-induced heating and amounts to merely a factor of
two increase only in the particle's time-averaged kinetic energy
\cite{cirac94}. 
These results are explicated in greater detail in Appendix
\ref{app:cooling-micromotion}.
(iii) We have numerically verified that both the expectation value for
the electron's motion as well as the corresponding fluctuations are
small compared to the SAW wavelength
$\lambda=2\pi/k$, i.e., $k\left\langle \hat{x}\right\rangle _{t}\ll1$
and $k\sigma_{x}\ll1$ with
$\sigma_{x}^{2}=\left\langle \hat{x}^{2}\right\rangle _{t}-\left\langle \hat{x}\right\rangle _{t}^{2}$,
thereby justifying our Lamb-Dicke approximation (with
$\cos\left(k\hat{x}\right)\approx\mathds1-\left(k^{2}/2\right)\hat{x}^{2}$)
self-consistently.

\textit{Self-consistency requirements}.\textemdash
Our theoretical framework is valid provided that the following
conditions are satisfied:
(i) First, the Markov approximation holds given that autocorrelations of
the bath (which typically decay on a timescale $\sim\hbar/k_{B}T$) decay
quasi instantaneously on the timescale of system correlations
$\sim\gamma^{-1}$ \cite{kohler97}. 
In principle, the damping rate $\gamma$ should be replaced by the
thermally enhanced rate
$\gamma_{\mathrm{eff}}=\gamma\left(\bar{n}_{\mathrm{th}}\left(\omega_{0}\right)+1\right)$; 
however, we will be interested mostly in the low-temperature,
pseudopotential regime where $\gamma_{\mathrm{eff}}\approx\gamma$. 
Thus, the Markov approximation yields the condition
$\hbar\gamma\ll k_{B}T$.
(ii) Second, the (weak-coupling) Born approximation holds provided that
the dissipative damping rate $\gamma$ is small compared to the relevant
system's transition frequencies, yielding the requirement
$\gamma\ll\omega_{0}$.
In the low-$q$ limit, taking conditions (i) and (ii), together with the
prerequisite for efficient ground-state cooling,
$k_{B}T\ll\hbar\omega_{0}$, yields the chain of inequalities
$\hbar\gamma\ll k_{B}T\ll\hbar\omega_{0}$.
In this regime, the weak-coupling Born approximation
$\left(\gamma\ll\omega_{0}\right)$ is satisfied very well.
(iii) Third, the characteristic separation of timescales between the
(slow) secular motion and the (fast) micromotion, with
$\omega_{0}=\varepsilon\omega$ and $\varepsilon=q/2\sqrt{2}\ll1$, gives
the requirement $\omega_{0} \ll \omega$.
(iv) Fourth, the energy scale $\hbar \omega$ has to be much smaller than
$E_{S}$ in order to ensure the existence of at least one bound state per
lattice site;  the latter can be estimated as
$n_{b}=V_{0}/\hbar \omega_{0}=\varepsilon E_{S}/\hbar \omega = (\varepsilon/2) m v_{s}/\hbar k$, 
leading to $\hbar \omega \ll E_{S}$ in the regime
$\varepsilon \ll 1$, $n_{b} \gtrsim 1$. 
Note that the existence of at least one bound state per lattice site
($n_{b} \gtrsim 1$) may always be fulfilled by choosing the lattice
spacing $a=\pi/k$ sufficiently large, at the expense of more severe
temperature requirements for ground state cooling and smaller energy
scales in the emerging Hubbard model (see below).  
Finally, the parameter regime of interest can be condensed into one line
of inequalities as ($\hbar=1$)
\begin{equation}
\gamma\ll k_{B}T\ll \omega_{0}\ll \omega \ll E_{S}.\label{eq:parameter-regime-of-interest-1}
\end{equation}
Let us discuss the implications of
Eq.$\ $\eqref{eq:parameter-regime-of-interest-1} in some more detail:
(i) In the parameter regime described by
$\ $Eq.$\ $\eqref{eq:parameter-regime-of-interest-1} the acoustic trap
is stable against thermal fluctuations, because $k_{B}T\ll V_{0}$ with
$V_{0}=\varepsilon^{2}E_{S}$; in other words,
$V_{0}=n_{b} \omega_{0} \gg k_{B}T$, if $\omega_{0} \gg k_{B}T$ and
$n_{b} \gtrsim 1$, as desired.
The condition $k_{B}T\ll \omega_{0}$, however, may be relaxed if
ground-state cooling is not necessarily required, akin to the physics of
optical tweezers. 
In this case, the less stringent condition $V_{0}\gg k_{B} T$ still
ensures a thermally stable trap.
(ii) The self-consistency requirement $\gamma \ll k_{B}T$ derives from
the Markov assumption of having  a short correlation time of the phonon
bath $\gamma \tau_{c} \ll 1$, with $\tau_{c}\sim 1/k_{B}T$. 
However, in the low-temperature regime, the correlation time $\tau_{c}$
may as well be set by the bandwidth of the bath $\Delta_{B}$ (that is,
the frequency range over which the bath at hand couples to the system),
rather than just temperature.
In that case, one may drop the condition $\gamma \ll k_{B}T$, leading to
a slightly refined regime of interest with
$\gamma, k_{B}T \ll \omega_{0}\ll \omega \ll E_{S}$, provided that the
Markov assumption $\gamma \tau_{c} \ll1$ is still satisfied with
$\tau_{c} \sim \Delta_{B}^{-1}$.
(iii) As a direct consequence of the presence of Mathieu-type instabilities, 
the proposed setup operates at relatively low
SAW-induced amplitudes set by the energy scale $E_{S}$, 
with the potential amplitude due to a single IDT given as
$V_{\mathrm{IDT}}=V_{\mathrm{SAW}}/2 = (q/2) E_{S} < E_{S}$.

Again, Eq.~\eqref{eq:parameter-regime-of-interest-1} underlines a
remarkably close connection to the established field of trapped ions,
where (as a direct consequence of Mathieu's equation, just as in our
setting) the inherent separation of timescales ($\omega_{0} \ll \omega$)
between (slow) secular motion and (fast) micromotion is well-known, 
albeit at very different energy scales with typical driving frequencies
$\omega/2\pi \sim 100\mathrm{kHz} - 100\mathrm{MHz}$ \cite{leibfried03}. 
Beyond this close analogy, our work identifies the importance of the
energy scale $E_{S}=(m/2)v_{s}^{2}$ in the proposed solid-state,
SAW-based setting, as displayed by
$\ $Eq.$\ $\eqref{eq:parameter-regime-of-interest-1}. 
Moreover, the first two inequalities in
$\ $Eq.$\ $\eqref{eq:parameter-regime-of-interest-1} derive directly
from the intrinsic solid-state cooling mechanism provided by the phonon
bath, whereas ions are typically cooled down to the motional ground
state using laser-cooling techniques that (as opposed to our solid-state
approach) explicitly involve the ion's internal level structure
\cite{leibfried03}.

In the following we will address the experimental implications of the
requirements listed in
$\ $Eq.$\ $\eqref{eq:parameter-regime-of-interest-1} for realistic
setups and show how some of the conditions may in fact be relaxed. 

\section{Implementation: How to Meet the Requirements \label{sec:implementation}}

Our previous conceptual analysis has revealed a specific set of
requirements [as summarized in
$\ $Eq.$\ $\eqref{eq:parameter-regime-of-interest-1}] which should be
fulfilled in order to ensure a faithful implementation of the proposed
AL setup in an actual experiment.
In the following we discuss several practical strategies in order to
meet these conditions.
Thereafter, we address several practical considerations 
which might be relevant under realistic experimental conditions.

\textit{Requirements}.\textemdash
First, rough 
(potentially optimistic; see below)
estimates for the spontaneous emission rate of acoustic
phonons $\sim\gamma$ may be inferred from low-temperature experiments on
charge qubits in (GaAs) double quantum dots which indicate rates as low
as $\gamma/2\pi \gtrsim 20\mathrm{MHz}$ ($\hbar\gamma \gtrsim 0.1\mu$eV)
\cite{fujisawa02,fujisawa98,hayashi03,petta04}.
We consider this estimate for the relaxation rate $\sim\gamma$ to be an 
optimistic, but still
adequate ballpark value for our SAW-induced acoustic traps, because the
typical
(i) temperatures ($T\sim\left(20-100\right)\mathrm{mK}$),
(ii) length-scales ($\sim300\mathrm{nm}$ for the dot-to-dot distance),
(iii) transition frequencies ($\sim\mathrm{GHz}$ in
Ref.\cite{hayashi03}), and
(iv) host materials (GaAs) studied in
Refs.\cite{fujisawa02,fujisawa98,hayashi03,petta04} are all compatible
with our setup. 
Furthermore, in Ref.\cite{hayashi03} a ohmic spectral density has been
assumed (just like in our theoretical model discussed above) in order to
fit the experimental data with the (thermally enhanced) decoherence rate
$\gamma_{\mathrm{eff}}=\gamma\left(2\bar{n}_{\mathrm{th}}(\omega_{\text{0}})+1\right)$,
yielding $\gamma=\zeta \omega_{0}$ with the fit parameter
$\zeta=(\pi/4)\times0.03\sim2.35\times10^{-2}$.
Second, we consider typical dilution-fridge temperatures in the range of
$T\sim\left(10-100\right)\mathrm{mK}$ (corresponding to
$k_{B}T\sim\left(1-10\right)\mu\mathrm{eV}$) \cite{barthelemy13}.
For $\gamma/2\pi \approx 20\mathrm{MHz}$ the first inequality in
Eq.\eqref{eq:parameter-regime-of-interest-1} is then safely satisfied
even for the lowest temperatures under consideration 
($k_{B}\cdot10\mathrm{mK}/2\pi\sim200\mathrm{MHz}$). 
Still, since $\gamma$ varies significantly with both energy and length
scales, phonon relaxation rates of $\gamma/2\pi \approx 20\mathrm{MHz}$
for GaAs-based systems may be overly optimistic. 
In this case, operation at higher temperatures [in order to satisfy
Eq.\eqref{eq:parameter-regime-of-interest-1}] may still be avoided by
employing (for example) phonon band gaps as discussed in
Ref.\cite{fujisawa98} or different materials such as silicon
\cite{kornich14,buy12} where the corresponding phonon-induced relaxation
rates are much smaller \cite{wang13}, as a consequence of a much smaller
electron-phonon coupling strength. 
All other things being equal, the SAW-induced potential depth
$V_{\mathrm{SAW}}$ will be reduced as well in a silicon-based setup,
which, however, can be compensated by simply applying a larger RF power
to the IDTs. 
Lastly, recall that the spontaneous emission rate $\gamma$ may be as
large as $\gamma \approx k_{B}T$ and still be fully compatible with the
desired regime of interest, if the correlation time of the phonon bath
is set by (for example) the bandwidth $\Delta_{B}$ rather than
temperature.
Third, for high SAW frequencies $\omega/2\pi\approx25\mathrm{GHz}$
\cite{kukushkin04}, the energy $\hbar\omega\approx100\mu\mathrm{eV}$
yields a trapping frequency $\hbar\omega_{0}\lesssim20\mu\mathrm{eV}$
($q^{2}\ll1$).
Altogether, we thus conclude that
Eq.\eqref{eq:parameter-regime-of-interest-1} can be satisfied with
state-of-the art experimental setups, provided that the
material-specific energy scale $E_{S}$ is much larger than 
$\hbar\omega\approx100\mu\mathrm{eV}$. 
For electrons in standard GaAs and the \textit{lowest} Rayleigh mode,
however, we find $E_{S}\approx2\mu$eV. 
In the following, we identify three potential, complementary strategies
to solve this problem.

\textit{(1) Material engineering}.\textemdash
Our first approach involves sophisticated material engineering, with the
aim to crank up the energy scale $E_{S}$.
Here, we can identify three general, complementary strategies to
increase the sound energy
{[}cf.$\ $Eq.$\ $\eqref{eq:energy-scale-Es}{]}.
(i) First, the effective mass $m$ crucially depends on both (a) the type
of particle and (b) the host material: 
(a) heavy holes or composite quasi-particles such as trions (also known
as charged excitons) typically feature much higher effective masses than
electrons in GaAs. 
(b) Compared to standard GaAs, where the effective electron mass is
$m\approx0.067m_{0}$ ($m_{0}$ refers to the free electron mass), in
Si/SiGe structures $m\approx0.2m_{0}$, while for electrons (heavy holes)
in AlN $m\approx0.33m_{0}$ ($m_{hh}=3.89m_{0}$).
(ii) Second, following common practice in the quest for SAW devices
operating at ultra-high frequencies
\cite{rodriguez12, benetti05, assouar07}, $v_{s}$ can be effectively
increased by employing a specialized heterostructure involving for
example diamond (which features the highest speed of sound).
(iii) Third, the speed of sound $v_{s,\alpha}=\omega_{\alpha}/k$ can be
enhanced even further by exciting higher-order Rayleigh modes $(\alpha>1)$
in the sample at the same wavelength \cite{morgan07}.
In particular, layered half-space structures (such as AlN/diamond, with
$h$ denoting the thickness of the piezoelectric AlN layer) support
so-called pseudo-surface acoustic waves (PSAWs) propagating with
exponential attenuation due to wave energy leakage into the bulk, in
contrast to regular (undamped) SAWs
\cite{morgan07,glushkov12,benetti05b}.
As shown both theoretically and experimentally
\cite{glushkov12,benetti05b}, this leakage loss can, however, become
vanishingly small for certain \textit{magic}
film-thickness-to-wavelength ratios $h/\lambda$, such that for all
practical purposes this PSAW mode can be seen as a true SAW mode which
propagates with negligible attenuation.
While SAWs by definition may not exceed the shear wave velocity $c_{s}$ 
($c_{s} \approx 12.32\mathrm{km/s}$ for diamond) in the lower
half-space, PSAW velocities can be significantly larger than $c_{s}$ and reach
values of up to $v_{s} \approx 18\mathrm{km/s}$
\cite{glushkov12,benetti05b}, that is about 40\% higher than those of
regular SAWs \cite{benetti05b} and about a factor of $\sim3.2$ higher as
compared to the lowest Rayleigh mode for a homogeneous AlN half-space
where $v_{s}\approx 5.6\mathrm{km/s}$ \cite{glushkov12}.
Lastly, even higher velocities may be achieved if leakage losses into
the bulk are suppressed when using freely suspended two-dimensional
electron gases \cite{blick98} rather than a (quasi) semi-infinite
half-space.

\begin{figure}[t]
\includegraphics[width=1\columnwidth]{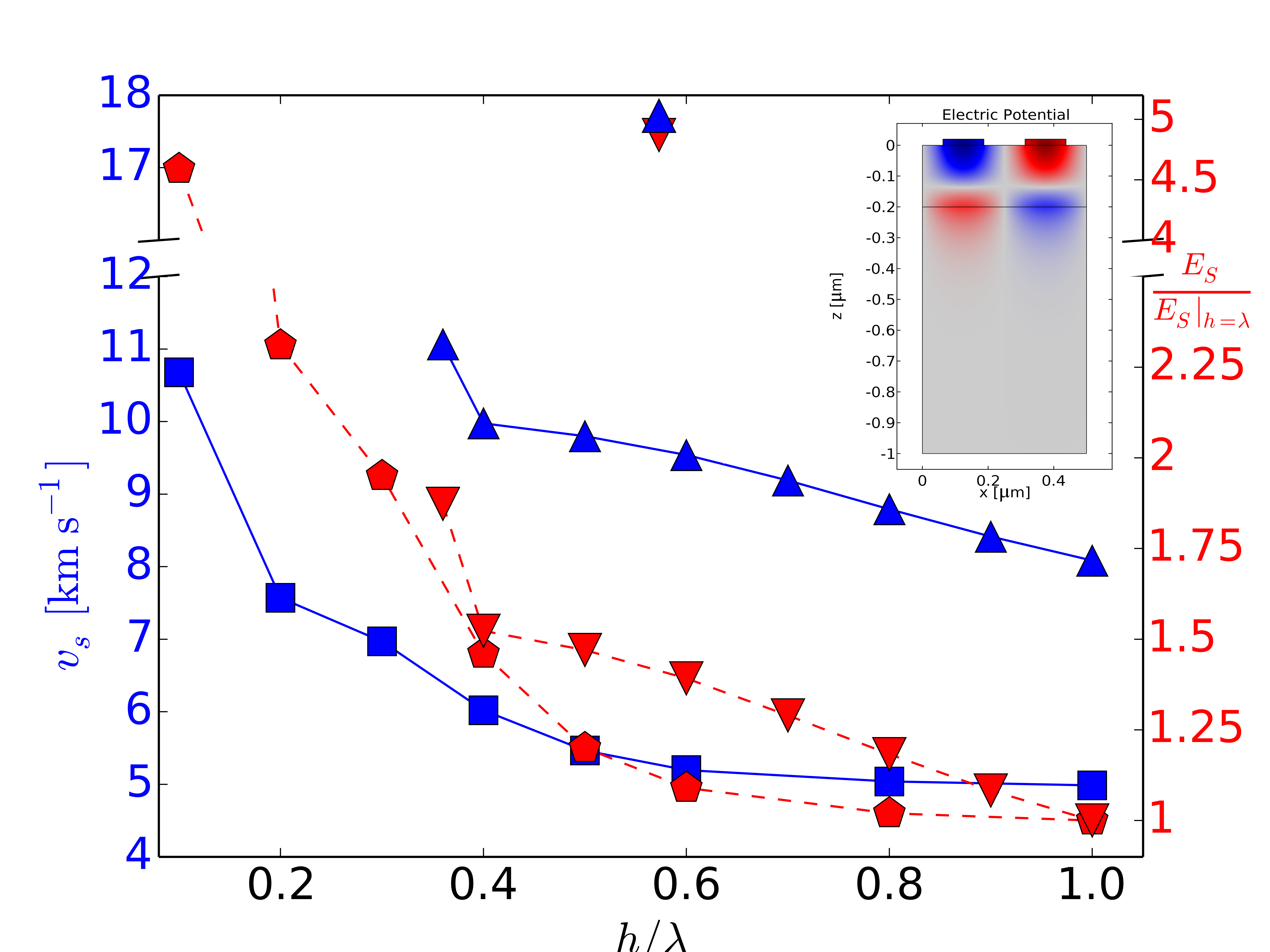}
\caption{\label{fig:comsol-1}(color online).
Speed of sound $v_{s}$ (left axis) and kinetic sound energy $E_{S}$
normalized to its value at $h = \lambda$ (right axis) in layered
heterostructures made of gallium arsenide (aluminium nitride) and
diamond.
All results are given as a function of $h$, which denotes the thickness
of the GaAs (AlN) layer.
Results for the second SAW modes and heavy holes are shown.
Squares and pentagons (triangles) denote the numerical results for a
GaAs/diamond (AlN/diamond) heterostructure.
$E_S (h = \lambda) \approx 32 \mu\mathrm{eV}$ for GaAs/diamond
($\approx 205\mu\mathrm{eV}$ for AlN/diamond).
The data points are connected by lines to guide the eye.
The isolated data points at $h \approx 0.57 \lambda$ denote a ultra-high
velocity PSAW mode in AlN/diamond (cf. Ref.\cite{glushkov12}).
Inset: Distribution of the piezoelectric potential at $f=12.2$ GHz of a
second SAW mode for a layer thickness of $h=0.2\mu$m in a GaAs/diamond
heterostructure.
The IDT finger spacing, hence the SAW wavelength, is set to be
$\lambda=500$ nm.
The results were obtained with the software package \textit{COMSOL}.}
\end{figure}

\begin{table}
\begin{tabular}{l||c|c|c|}
setup & $m/m_{0}$ & $v_{s}[\mathrm{km/s}]$ & $E_{S}[\mu\mathrm{eV}]$\tabularnewline
\hline 
\hline 
electrons in GaAs$^{*}$ & 0.067 & $\sim3$ & $\sim1.7$\tabularnewline
\hline 
heavy holes in GaAs$^{**}$ & 0.45 & $\sim(12-18)$ & $\sim184-415$\tabularnewline
\hline 
electrons in Si$^{**}$ & 0.2 & $\sim(12-18)$ & $\sim82-184$\tabularnewline
\hline 
holes in GaN$^{**}$ & 1.1 & $\sim(12-18)$ & $\sim450-1010$\tabularnewline
\hline 
electrons in $\mathrm{MoS}_{2}$$^{**}$ & 0.67 & $\sim(12-18)$ & $\sim274-617$\tabularnewline 
\hline 
trions in $\mathrm{MoS}_{2}$$^{**}$ & 1.9 & $\sim(12-18)$ & $\sim794-1787$\tabularnewline
\hline 
\end{tabular}
\caption{\label{tab:Es}
Estimates for the energy scale $E_{S}$ for different physical setups.
Examples marked with $^{*}$ refer to the lowest SAW mode in GaAs whereas
those marked with $^{**}$ refer to relatively fast (diamond-boosted)
values of $v_{s}$ in diamond-based heterostructures featuring
high-frequency SAW and PSAW modes as investigated in
\cite{glushkov12,benetti05b}.
Further details are given in the text. }
\end{table}

We have verified these considerations using numerical finite-element
calculations, performed with the software package \textit{COMSOL}
\cite{comsol} for GaAs/diamond (AlN/diamond) heterostructures; our
simulations indeed show that the effective speed of sound can be
significantly scaled up in comparison with the standard values in
GaAs (AlN) \cite{footnote-heterostructure}. 
In Fig.$\ $\ref{fig:comsol-1}, the behaviour of $v_{s}$ as a function of
the width $h$ of the GaAs (AlN) layer is displayed.
The results show both the second Rayleigh SAW modes in GaAs/diamond and
AlN/diamond, respectively, as well as one particular PSAW mode (as
identified previously in Ref.\cite{glushkov12}). 
For large $h$, the second Rayleigh SAW modes coincide with the
corresponding second modes in the raw materials GaAs and AlN (without a
diamond layer), as expected.
On the other hand, in the limit of comparatively small
$h_{\text{}}\approx(50-200)$ nm, the SAW velocities are significantly
larger compared to the first and second Rayleigh modes in pure GaAs
(AlN), while for the PSAW mode
$v_{s} \approx 18\mathrm{km/s}$ at $h/\lambda \approx 0.57$. 
Moreover, in the case of piezoelectric coupling, the electric potential
which accompanies the SAW has to be non-zero at the 2DEG which is
located somewhere in the center of the top GaAs (AlN) layer.
As shown in Fig. \ref{fig:comsol-1}, such configurations do exist in
GaAs/diamond (AlN/diamond) heterostructures, while reaching the
parameter regime
$k_{B}T=1\mu\mathrm{eV} \lesssim (10^{-3}-10^{-2})E_{S}$. 
Hence, when suitably combining strategies (i)-(iii), we predict the
feasibility of reaching $E_{S}\gtrsim1$ meV, which is sufficiently large
to safely satisfy condition \eqref{eq:parameter-regime-of-interest-1},
as desired.
Consider for example a two-dimensional hole gas at a AlN/GaN interface
on top of diamond; here, the effective heavy-hole mass of GaN amounts to
$m\approx1.1m_{0}$. 
When driving the PSAW mode identified in Fig.~\ref{fig:comsol-1}, we
find $E_{S}\approx 1.0\mathrm{meV}$. 
Alternatively, we may consider monolayer transition metal
dichalcogenides (TMDCs) such as $\mathrm{MoS}_{2}$ or $\mathrm{WSe}_{2}$,
on top of some high-speed material such as diamond.  
While all TMDCs are piezoelectric due to the lack of inversion symmetry
\cite{novoselov16}, some of them show relatively large effective masses; 
for example, the effective electron and hole mass in $\mathrm{MoS}_{2}$
amount to approximately $m\approx0.67m_{0}$ and $m\approx0.6m_{0}$,
respectively \cite{kormanyos15, eknapakul14}.
Then, for electrons (charged trions) in $\mathrm{MoS}_{2}$ with effective mass
$m\approx0.67m_{0}$ ($m\approx1.9m_{0}$) \cite{eknapakul14}, as experimentally investigated
for example in Refs.\cite{mak13,ross13}, and a diamond-boosted
speed-of-sound $v_{s}\approx 18\mathrm{km/s}$, we estimate $E_{S}$ to be
as large as $E_{S}\approx 617\mu \mathrm{eV}$ ($E_{S}\approx 1.78\mathrm{meV}$). 
Further estimates of this type for different physical setups are
summarized in Tab.\ref{tab:Es}.
Here, we have covered the most relevant material properties for the
implementation of the proposed AL setups only, whereas the interplay of
different material-design strategies (i)-(iii), leads to an intricate
problem involving various parameters (such as piezoelectric properties
and the electron mobility), which we cannot cover in its full depth
within the scope of this work.

While this material-engineering based approach is fully compatible with
our general theoretical framework, as described in Section
\ref{sec:theory}, in the following we present two additional schemes that
allow for thermally stable trapping, at potentially higher temperatures
than what we have found so far, but at the expense of a more involved
theoretical description [which, however, is not necessarily restricted
to the parameter regime given in
Eq.\eqref{eq:parameter-regime-of-interest-1}]; here, similar to Section
\ref{sec:theory}, we first present a classical analysis of the dynamics, 
whereas a detailed, quantum-mechanical analysis thereof goes beyond the
scope of this work and will be subject to future research.

\textit{(2) Exotic stability regions}.\textemdash
In the context of ion traps where stability is governed by the Mathieu
equation [cf. Eq.(\ref{eq:mathieu})], ion motion is stable in the primary
stability region $(a_{\mathrm{dc}}=0,\ 0<q<0.908)$ and then becomes
unstable as $q$ is increased  \cite{leibfried03}. 
Stable motion, however, reoccurs at higher $q$ values which we refer to
as exotic stability regions in the following; 
these exotic stability regions were studied to some extent in the context
of ion traps \cite{dawson1,dawson2}. 
Here, we propose, as a second strategy to meet the self-consistency
requirements, to extend the previously established classical stability
analysis to the next higher-lying 
$(a_{\mathrm{dc}}=0,\ 7.5 \lesssim q \lesssim 7.6)$ stability region of
the Mathieu equation.
As evidenced in Fig.\ref{fig:stability}(d), in this high-$q$ regime, a
separation between secular and fast (micro-)motion is no longer possible.
However, while the theoretical description of the dynamics becomes more
involved, still the particles are found to be dynamically
\textit{trapped}, already at temperatures much higher than what we found
in the low-$q$ regime.
While $k_{B}T\lesssim0.03E_{S}$ for small $q$, in the high-$q$ regime
(with $7.5<q<7.6$) thermal stability sets in already at
$k_{B}T\lesssim0.15E_{S}$, thus alleviating temperature requirements by
about an order of magnitude, cf. Fig.\ref{fig:stability}(c).

\textit{(3) Optimized driving schemes}.\textemdash
As a third strategy, we suggest to utilize polychromatic driving
schemes, rather than the simple monochromatic driving considered so far. 
Recently, it has been experimentally demonstrated that such
polychromatic drivings can eventuate arbitrary SAW wavefronts
\cite{schuelein15}, thus allowing us to consider more general equations
of motion of the form $\ddot{x}+f(\tau)x=0$, with some particular time
dependence $f(\tau)$.
For example, instead of the Mathieu equation for which
$f(\tau)=2q\cos(2\tau)$ (no dc contribution), a simple two-tone driving
scheme can be used to expand the stability regions as previously
suggested in Ref.\cite{possa16}.
Our numerical studies suggest that the superposition of higher harmonics
in the form of $f(\tau)=2q[c_{1}\cos(2\tau)+c_{2}\cos(4\tau)+...]$ may
already enhance the robustness of the stability region in Fig.
\ref{fig:stability}(a) against temperature by a factor of two, as
compared to the standard Mathieu equation. 

\textit{Technical considerations}.\textemdash
We now address several technical considerations which might be relevant
for a faithful experimental realization of our proposal: 
(i) Since the potential amplitude due to a single IDT is limited by Mathieu-type
stability arguments as $V_{\mathrm{IDT}}=V_{\mathrm{SAW}}/2=(q/2)E_{S}\lesssim0.5\mathrm{meV}$ \cite{footnote-amplitude},
the proposed setup operates at SAW-induced amplitudes that are about
two orders of magnitude smaller than what is common for SAW-induced
electron transport experiments (where typically $V_{\mathrm{IDT}}\approx40\mathrm{meV}$ \cite{furuta04,hermelin11}).
Note that this comparatively low driving amplitude amounts to a fraction
of typical quantum dot charging energies. Today, quantum dots
are routinely pulsed with similarly high amplitudes, and yet excellent
charge and spin coherence is seen in experiments \cite{baart17,martins16,reed16}.
(ii) In a similar vein, as a direct consequence of the low-amplitude external drive, 
potential microwave-induced heating effects of the sample should be small. 
Furthermore, undesired heating may be suppressed efficiently by placing the IDTs
very far away from the center of the trap, without losing acoustic power, thereby avoiding 
local heat dissipation near the center of the trap due to the applied RF power;
for further details we refer to Appendix \ref{sec:case-study}.
(iii) Minimization of crosstalk-related effects can be accomplished based on various techniques \cite{campbell89}:
these can involve, for example, very careful choice of metal-packaging
structure and dimensions, the judicious placement of ground connections
to avoid ground loop effects, and the placement of thin metal-film
ground strips between the IDTs.
Moreover, because of the vast difference between
the speed of light $\left(c\approx10^{8}\mathrm{m/s}\right)$ and
the speed of sound $\left(v_{s}\approx10^{4}\mathrm{m/s}\right)$,
for a given frequency the wavelength associated with the EM crosstalk
is about four orders of magnitude larger than the SAW wavelength (even
when accounting for the refractive index of the specific material),
and therefore practically flat on the relevant lengthscale of a few
lattice sites; for $\omega/2\pi\approx30\mathrm{GHz}$, the wavelength
is in the millimeter range, i.e., much larger than the acoustic lattice
spacing $a=v_{s}/\left(\omega/\pi\right)\approx170\mathrm{nm}$. 


\section{Applications\label{sec:Many-Body-Physics}}

The possibility to acoustically trap charged particles in a
semiconductor environment should open up many experimental
possibilities, well beyond the scope of this work. 
Here, we briefly describe just two potential exemplary applications; 
see also our discussion in the final section of this paper.

\subsection{Mobile Acoustic Quantum Dots}

\begin{figure}
\includegraphics[width=0.9\columnwidth]{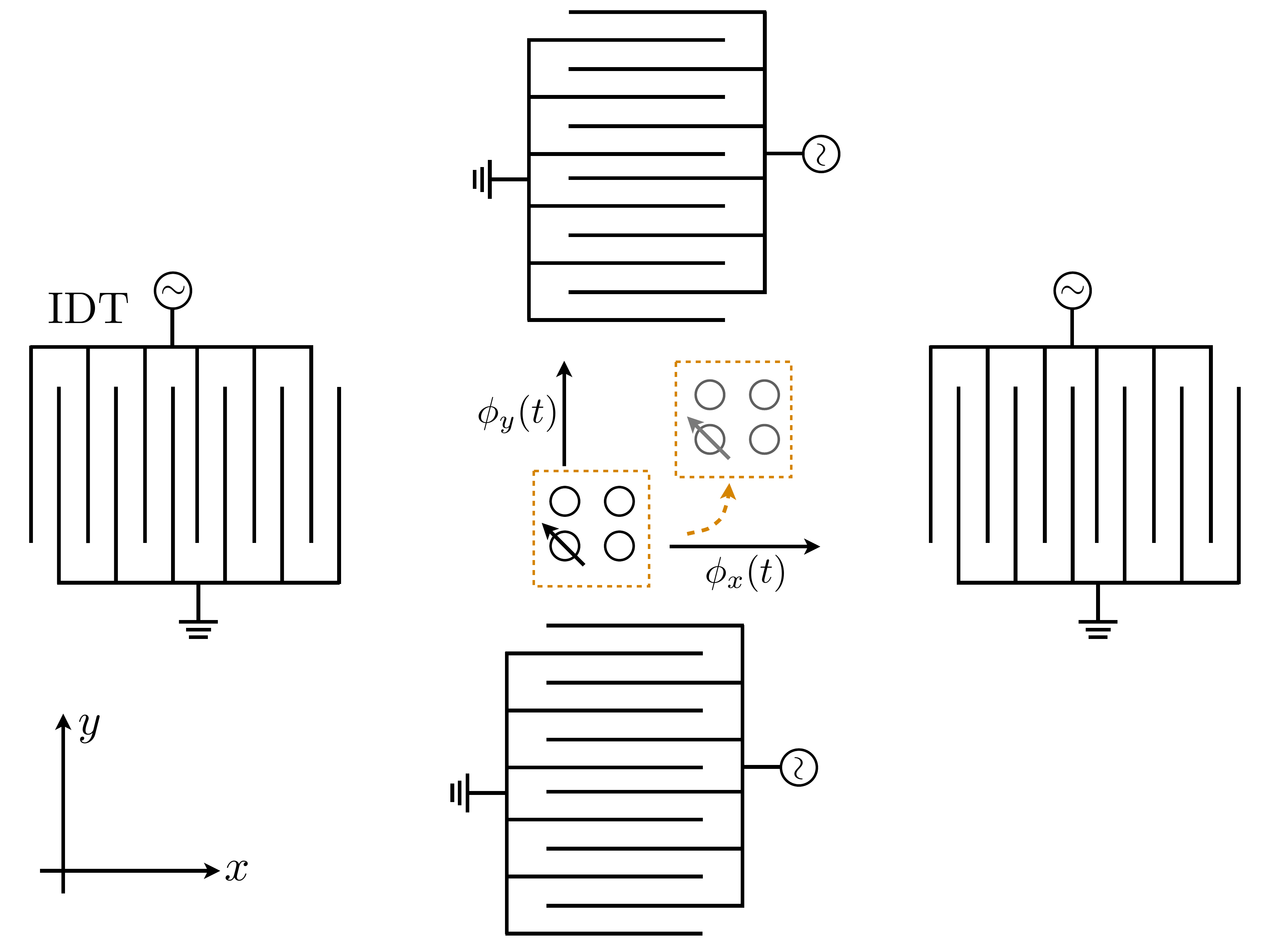}
\caption{\label{fig:mobile-AL}(color online). 
Schematic illustration (top view; not to scale) for a quasi-stationary
two-dimensional AL which can be controllably displaced in both $x$- and
$y$-direction by adiabatically tuning the phases applied to the IDTs.
The dashed (orange) box highlights a small sub-lattice consisting of
just four lattice sites, before and after the adiabatic ramp.}
\end{figure}

\textit{Mobile acoustic quantum dots}.\textemdash
By adiabatically tuning the phases applied to the IDTs one may displace
the AL in both the $x$- and $y$-direction, thereby creating
\textit{mobile} acoustic quantum dots, with the possibility to transfer
in this way quantum information stored in the spin degree of freedom of
the particle; for a schematic illustration compare
Fig.~\ref{fig:mobile-AL}. 
Here, in contrast to standard SAW-based mobile quantum dots
\cite{hermelin11,mcNeill11}, the speed $v_{\mathrm{eff}}$ at which the
trapped particles can be moved around between different locations in the
2DEG would not simply be set by the SAW's speed of sound $v_{s}$, but
could rather be controlled \textit{in situ} by the time derivative of
the phases applied to the IDTs, with an upper bound roughly given by
the adiabaticity condition ($\varepsilon_{\mathrm{ad}}\ll1$) as 
$v_{\mathrm{eff}}= \varepsilon_{\mathrm{ad}} a(\omega_{0}/2\pi) \lesssim 100\mathrm{m/s}$.  
Apart from thermal fluctuations, the trapping lifetime in such a mobile
quantum dot will be limited by tunnel-coupling to neighbouring mobile
quantum dots inside the AL (while the spin lifetime remains unaffected
for spin-coherent tunneling); as shown in more detail below, this
coupling can be suppressed controllably by going to a larger SAW
wavelength $\lambda$, at the expense of more stringent ground-state
cooling requirements as the level spacing $\omega_{0}$ decreases.
For a mobile AL with near unit filling, however, tunnelling is largely
suppressed due to Coulomb blockade effects and the (spin) dynamics is
governed by the next lower energy scale (the exchange coupling), as
discussed and quantified next.

\subsection{Towards Many-Body Physics}

\textit{Many-Body physics}.\textemdash While our previous discussion has
exclusively focused on dynamically trapping and cooling \textit{single}
particles in SAW-induced potentials, in our second example we provide a
simple characterization of our setup for the potential investigation of
quantum \textit{many-body} systems. We show that (at dilution fridge
temperatures) our system can be naturally described by an extended
Anderson-Hubbard model, with the ultimate prospect of entering the low
temperature, strong interaction regime where $k_{B}T\ll t<U$; here $t$
and $U$ refer to the standard hopping and interaction parameters of the
Hubbard model, as specified below. We provide estimates for these
quantities in terms of the relevant parameters characterizing the AL,
and show how they can be engineered and (dynamically) tuned.
For this analysis, again we restrict ourselves to the pseudopotential
regime ($\omega_{0} \ll \omega$) where the effects of the fast,
small-amplitude micromotion on the Hubbard parameters $t,U$ can be
neglected.
Thereafter, we discuss several approaches that may be used in order to
detect and accurately probe the resulting quantum phases of matter.

\textit{Estimates for Hubbard parameters}.\textemdash
Consider an ensemble of fermionic charged particles inside a periodic
one or two-dimensional AL, with roughly one particle per site
(corresponding to electron densities $\sim10^{10}\mathrm{cm}^{-2}$ for a
two-dimensional AL with $a\sim100\mathrm{nm}$). 
If all energy scales involved in the system dynamics are small compared
to the excitation energy to the second band $\sim\hbar\omega_{0}$
(for example $k_{B}T\ll\hbar\omega_{0}$, as required for ground state
cooling), the electrons will be confined to the lowest Bloch band
of the AL, and the system can effectively be described by the extended
Anderson-Hubbard Hamiltonian
\cite{hofstetter02,byrnes07,jaksch98,anderson75}
\begin{eqnarray}
H_{\mathrm{AFH}} & = &
-t\sum_{\left<i,j\right>,\sigma}\left(c_{i,\sigma}^{\dagger}c_{j,\sigma}
+\mathrm{h.c.}\right)+\sum_{i}\mu_{i}n_{i}\nonumber \\
 &  & +\sum_{\sigma,\sigma'}\sum_{ijkl}U_{ijkl}
 c_{i,\sigma'}^{\dagger}c_{j,\sigma}^{\dagger}
 c_{k,\sigma}c_{l,\sigma'},\label{eq:Anderson-Hubbard-Hamiltonian}
\end{eqnarray}
where the fermionic operator $c_{i,\sigma}(c_{i,\sigma}^{\dagger})$
annihilates (creates) an electron with spin $\sigma=\uparrow,\downarrow$
at site $i$; $n_{i,\sigma}=c_{i,\sigma}^{\dagger}c_{i,\sigma}$ and
$n_{i}=n_{i,\uparrow}+n_{i,\downarrow}$ refer to the spin-resolved
and total occupation number operators, respectively. In
Eq.(\ref{eq:Anderson-Hubbard-Hamiltonian}) we have retained the nearest
neighbour hopping term only, as specified by a tunneling amplitude $t$,
but accounted for the full effect of the repulsive (long-range) Coulomb
interactions $\sim U_{ijkl}$.
The remaining (second) term, with a variable on-site energy $\mu_{i}$,
acts like a spatially varying chemical potential and describes potential
disorder effects (as discussed in more detail below). In the limit
of homogeneous on-site energies with $\mu_{i}=\mathrm{const.}$,
Eq.(\ref{eq:Anderson-Hubbard-Hamiltonian}) reduces to the minimal
Hubbard model, if all but the largest on-site interaction terms are
neglected (with $U_{iiii}=U$ in standard notation).
In the limit $V_{0}\gg E_{R}$ (where $E_{R}=\hbar^{2}k^{2}/2m$ is the
recoil energy), the tunneling rate $t$ is given by
$t/E_{R}\approx\left(4/\sqrt{\pi}\right)\left(V_{0}/E_{R}\right)^{3/4}\exp\left[-2\sqrt{V_{0}/E_{R}}\right]$
\cite{bloch08}, setting the upper limit $t<E_{R}$. 
In terms of the relevant AL parameters, this relation can be rewritten
as
\begin{equation}
t/E_{S}\approx\left(2\sqrt{2\pi n_{b}}\right)^{-1}q^{2}\exp\left[-4n_{b}\right],
\end{equation}
showing that the tunneling rate $t\sim q^{2}$ can be tuned via the
stability parameter $q$, reaching at maximum
$t/E_{S}\lesssim3\times10^{-3}$ within the lowest stability region for
(fixed) $n_{b}\gtrsim1$; here, the existence of at least one bound state
$\left(n_{b}\gtrsim1\right)$ ensures both
$V_{0}/E_{R}=4n_{b}^{2}\gtrsim4$ and $t\ll\hbar\omega_{0}$, as required.
Therefore, with $E_{R}/E_{S}=\varepsilon^{2}/(4n_{b}^{2})$, we find
$t<E_{R}\ll E_{S}$ such that realistically $E_{S}\gg300\mu\mathrm{eV}$
is required in order to access the coherent many-body regime where
$t\gg k_{B}T$, at dilution fridge temperatures. 
Making use of the strategies outlined in the previous section, this
regime seems to lie within reach of state-of-the-art experimental
capabilities. 
The order of magnitude for the Coulomb integral $U_{ijkl}$ can be
roughly estimated as $U_{ijkl}\sim e^{2}/4\pi\epsilon a$ (where
$\epsilon$ denotes the effective dielectric constant of the material).
Since $t\sim E_{R}\sim a^{-2}$ and $U_{ijkl}\sim1/a$, the relative
importance of the hopping term $\sim t$ as compared to the Coulomb
interactions can be conveniently controlled via the SAW frequency
$\omega=\pi v_{s}/a$ \cite{byrnes07}.
Taking (for example) $a\sim300\mathrm{nm}$, this rough estimate yields
$U_{ijkl}\sim380\mu\mathrm{eV}$ (for GaAs, where
$\epsilon\approx12.5\epsilon_{0}$), which exceeds any realistic hopping
amplitude $t$ by far, but also violates the assumptions underlying
the model Hamiltonian (\ref{eq:Anderson-Hubbard-Hamiltonian}).
To enter a parameter regime where the simplified toy model
(\ref{eq:Anderson-Hubbard-Hamiltonian}) becomes applicable, special
heterostructures with a metallic screening layer close to the 2DEG may
be employed, while in a similar vein the thickness of the spacer layer
(separating the 2DEG from the $\delta$-doping layer) may also be reduced
in favour of increased screening effects \cite{barthelemy13,byrnes07}.
In this scenario, a simple image-charge based estimate shows that the
Coulomb interaction is reduced by a factor
$f_{\mathrm{scr}}\approx1-[1+4\left(d/a\right)^{2}]^{-1/2}$ (where $d$
refers to the distance between the 2DEG and the metallic screening
plate), while retaining its $\sim1/r$ scaling \cite{byrnes07}.
Accordingly, the estimate quoted above reduces from
$\sim380\mu\mathrm{eV}$ for $d\rightarrow\infty$ down to
$\sim50\mu\mathrm{eV}$ for $d\sim0.3a\sim90\mathrm{nm}$.
As discussed in more detail below, this approach does not only allow
for tuning the strength of the Coulomb interaction (albeit not
\textit{in situ}), but at the same time reduces the detrimental effects
due to background impurities \cite{barthelemy13}.
In a regime where the latter is negligible, the next lower energy scale
is set by the exchange coupling $J=4t^{2}/U$, which describes effective
spin-spin interactions via virtual hopping processes in the regime
$U\gg t$.
With the Coulomb interaction reduced to $U\approx10t$, the regime
$t\gg k_{B}T$ (and therefore $J\approx t/2\gg k_{B}T$) should then give
access to experimental studies of quantum magnetism \cite{bloch08}.
For a comprehensive overview of the key quantities of our analysis and
self-consistent estimates thereof we refer to Appendix
\ref{sec:case-study}.
In this Appendix we also discuss relevant electron spin decoherence effects 
which may compete with the observation of coherent spin physics. 

\textit{Detection schemes}.\textemdash In order to measure the resulting
collective many body state in an actual experiment, several approaches
may be available: 
(i) First, the electron excitation spectrum could
be probed using inelastic light scattering, as has been done
experimentally in a closely related setup (based on electrons confined
in etched pillars in a high-quality GaAs quantum well with mobility
$\mu \sim 3\times 10^{6}\mathrm{cm^{2}/Vs}$) in Ref.\cite{singha11}.
(ii) Second, transport measurements, in which a small dc voltage
$V_{\mathrm{dc}}$ is applied across the AL, should carry signatures of
the phase of the Hubbard model in the resulting dc current; compare for
example Refs.\cite{byrnes07,schlosser96,albrecht01}. 
The corresponding dc current $I_{\mathrm{dc}}$ will be blocked in the 
Mott-insulator regime, whereas Ohm's law
$I_{\mathrm{dc}} \propto V_{\mathrm{dc}}$ should hold in a metallic
phase \cite{byrnes07,hensgens17}.
(iii) Third, charge-imaging methods could also be used to demonstrate
regular carrier localization in the acoustic lattice, somewhat similar
to the detection of single electrons trapped by impurity centres
\cite{kuljanishvili08,martin04}. 
(iv) Fourth, capacitance spectroscopy techniques (as demonstrated for
example in Ref.\cite{dial07}) could be used in order to measure the
density of states by detecting the ability to tunnel in from a
back-plane.
(v) Fifth, optical readout of the charge- and spin-state could be achieved with
methods developed for self-assembled quantum dots \cite{gywat10}, 
in particular in TMDC-based setups \cite{novoselov16,mak13,ross13,srivastava15}. 
Similar to self-assembled quantum dots, our SAW-defined quantum dots and lattices trap both electrons
and holes at the same location and could thus support quantum-dot
excitons and trions. The charge- and spin-dependent interaction with
quasi-resonant light fields can be used for read-out via resonance
fluorescence \cite{vamivakas09} or the Kerr effect \cite{atatuere07, berezovsky06}. 
Moreover, it is conceivable that
related optical techniques for state preparation and spin rotation can
be adapted as well.
Note that due to the expected homogeneity of our SAW-generated lattice
sites, we also expect largely identical optical spectra across the
lattice which may facilitate global readout and collective optical effects.
(vi) Lastly, apart from these well-established measurement techniques, we
propose to perform local site-resolved detection by adiabatically
changing the phases at the IDTs $\phi\left(t\right)$ and then loading
one lattice site after the other (very much like in a CCD camera) into
nearby gate-defined quantum dots, where both the charge as well as the
spin degree of freedom could be measured via well-established
spin-to-charge conversion techniques \cite{baart16}.

\section{Effects of Disorder}

Disorder in the AL will affect
the (Anderson) Hubbard model, as described by the second term in
Eq.(\ref{eq:Anderson-Hubbard-Hamiltonian}), where (in the presence of
disorder) $\mu_{i}$ is essentially a randomly fluctuating variable.
In a semiconductor the dominant source of disorder is due to charged
impurities, which includes both (i) desired contributions (such as the
dopants used for forming the 2DEG) as well as (ii) undesired ones due to
bulk or surface impurities \cite{barthelemy13}.
While the dominant source of disorder (i) due to remote donor scattering
can be largely removed in structures with a relatively large spacer
thickness $\sim85\mathrm{nm}$ \cite{barthelemy13,byrnes08}, the second
one (ii) has been identified as the main mechanism limiting the mobility
$\mu$ in ultra-clean 2DEGs \cite{footnote-mobility, umansky97}.
Still, as experimentally demonstrated in Ref.\cite{umansky97},
mobilities exceeding $\sim10^{7}\mathrm{cm^{2}/Vs}$ can be realized for
dilution-fridge temperatures $T\sim100\mathrm{mK}$, resulting in a
mean-free-path
$l_{\mathrm{mfp}}=\mu v_{\mathrm{F}}m/e$ of up to
$l_{\mathrm{mfp}}\sim120\mu\mathrm{m}$ (here, $v_{\mathrm{F}}$ refers to
the Fermi velocity \cite{byrnes07}).
In the low-density regime of interest with
$n_{\mathrm{el}}\sim10^{10}\mathrm{cm}^{-2}$ (corresponding to
half-filling for a lattice spacing of $a\sim100\mathrm{nm}$) the
mean-free-path is expected to drop to
$l_{\mathrm{mfp}}\sim15\mu\mathrm{m}$ \cite{umansky97}, which is still
much larger than the lattice spacing $a\sim100\mathrm{nm}$.
To further compensate for residual disorder originating from background
impurities (ii) one may resort to special heterostructures with a
conducting backplane, as suggested in Ref.\cite{barthelemy13}.
Also, in periodic arrays of quantum dots signatures of Hofstadter's
butterfly \cite{hofstadter76} have been observed at high magnetic
fields \cite{albrecht01,schlosser96}, as a result of the interplay
between the periodic potential and quantized Hall orbitals, suggesting
that disorder from the substrate can in fact be sufficiently small
to investigate coherent lattice physics.
This discussion certainly provides the basis for some optimism, but a
dedicated research program (rather than just a literature survey) may be
required to fully understand and characterize the role of disorder in
this system; compare Ref.\cite{hensgens17} for recent efforts in this
direction based on gate-defined lattices in GaAs.
While the effect of disorder on the single-particle level is well
understood \cite{lagendijk09}, the intricate interplay between
interactions and disorder in the Hubbard model (as studied in
Refs.\cite{basko06,belitz94,byczuk05,fallani07}) yields a non-trivial
regime in its own right which may be explored systematically in the
proposed setup by deliberately controlling the amount of disorder.


\section{Summary \& Outlook }

In summary, we have proposed and analyzed the formation of an
all-solid-state acoustic lattice with a highly regular periodicity set
by the SAW wavelength (without any further gate patterning).
We have developed a theoretical framework reminiscent of trapped-ion
physics thus connecting two previously unrelated fields of research.
With this framework at our disposal, we have identified the relevant
figures of merit for this system and discussed potential experimental
platforms for a faithful implementation of such acoustic lattices, with
the ultimate potential to study yet unexplored parameter regimes,
thanks to specific system properties such as ultra-light particle masses,
intrinsic electron-phonon cooling and strong inter-particle interactions.
Here, let us emphasize again the flexibility (and generic nature) of the proposed
scheme: SAWs exist in many materials (semiconductor heterostructures,
TMDCs), can be endowed with a variety of accommpanying fields (depending
on the material used: strain, electric, magnetic) and superposed to
different standing wave patterns. Therefore the proposed scheme should be
applicable to a variety of different (quasi-) particles and allow to study
different lattice geometries.

Finally, we highlight possible directions of research going beyond
our present work:
(i) While we have focused on a simple square-lattice geometry, more
sophisticated lattice geometries might be explored, given the design
flexibilities associated with SAW devices \cite{morgan07}.
(ii) For simplicity, in this work we have disregarded the potential
presence of magnetic fields and/or spin-orbit effects, which stem from
the underlying material properties.
Therefore, without any further sophisticated engineering, these
additional ingredients could be readily implemented, giving rise to rich
phase diagrams and, for example, the formation of topological quantum
spin Hall states \cite{sushkov13}.
Finally, we may envisage several setups that are complementary to the
system studied in this work: 
(iii) \textit{Acoustic lattices for dipoles}:
Our ideas can be generalized towards an acoustic lattice for solid-state
dipoles (rather than charged particles), e.g., for  indirect excitons
which consist of electrons and holes from two different parallel quantum
well (QW) layers, thereby complementing previous experimental studies on
SAW-induced lattices for exciton-polaritons both in moving
\cite{cerda-mendez10} and standing-wave \cite{buller16} configurations
in the regime of many particles per lattice site.
As evidenced by several experiments (where the repulsive character of
the interaction shows up as a positive and monotonic line shift with
increasing density \cite{butov04}), indirect excitons behave as
effective dipoles perpendicular to the plane \cite{hammack06,hammack07}.
Because of the spatial separation between the electron and hole layers
in this coupled QW structure, the intrinsic radiative lifetimes of
optically active indirect excitons exceeds that of their direct
counterparts by orders of magnitude and can be in the range of several
microseconds \cite{hammack07}. 
In TMDC-based setups our approach may be used to \textit{dynamically} trap and to 
spatially and spectrally isolate single excitons, thereby complementing experiments
based on \textit{static} strain-engineering \cite{kumar15}. 
(iv) \textit{Acoustic lattices for ions}:
The electric potential (created and controlled at the surface) due to
standard IDTs extends into the material, but also into the vacuum above
the surface \cite{schuetz15}.
In principle, this should allow for the integration of our SAW-based
setup with ions above the surface that are exposed to this acoustically
induced electric potential, leading to new hybrid setups and
complementing other approaches towards regular, disorder-free surface
traps for ions in which the lattice spacing is simply set by the SAW
wavelength. 
With comparatively large parameter values for
$E_{S}$ ($\approx420\mathrm{meV}$ for Be ions on top of GaAs),
preliminary estimates show that a pseudopotential trap depth of several
$\sim\mathrm{meV}$ should be possible within the lowest stability region
(where $q^{2}\ll1$), provided that the ion can be stabilized in the
direction normal to the surface within the SAW wavelength.
(v) \textit{Magnetic lattices}:
While the acoustic lattice described above is based on coupling to the
particle's \textit{external} motional degree of freedom (as is the case
with Paul traps for ions), in closer analogy to optical lattices for
ultra-cold atoms, SAWs in piezo-magnetic materials such as Terfenol-D
\cite{weiler11,pang08,liu08} may be used in order to couple to the
particle's \textit{internal} spin degree of freedom, thereby inducing a
spatially inhomogeneous Stark shift on the electron's spin resonance
which will act as an external potential for the electron's motion
\cite{knoerzer17}.
In this setup, for a fixed detuning of the ESR driving frequency from
the Zeeman splitting, the effective trap depth can (in principle) be made
arbitrarily large, provided that sufficient SAW power is available.

In conclusion, this discussion indicates that by combining the control
and flexibility of SAWs with the rich variety of material properties of
heterostructures, the emerging field of quantum acoustics opens a large
number of further research directions with the ultimate goal of understanding
the behavior of correlated electrons in technologically relevant
materials and molecules and building a universal quantum simulator.

\begin{acknowledgments}
\textit{Acknowledgments}.\textemdash MJAS would like to thank the
Max-Kade foundation for financial support and MPQ for
hospitality.
MJAS, JK and JIC acknowledge support by the DFG within the Cluster of Excellence NIM. 
GG acknowledges support by the Spanish Ministerio de Economía y
Competitividad through the Project FIS2014-55987-P and thanks MPQ for
hospitality.
LMKV acknowledges support by a European Research Council Synergy grant
and the Netherlands Organization for Scientific Research (NWO).
Work at Harvard was supported by NSF, Center for Ultracold Atoms, CIQM, 
Vannevar Bush Fellowship, and AFOSR MURI.
MJAS and JK thank Alejandro Gonzalez-Tudela, Kristiaan De Greve, Hans Huebl, Eric Kessler,
Hubert Krenner, Florian Marquardt, Javier Sanchez-Yamagishi,
Paulo Santos, Robert Ukropec, Mathias Weiler, Dominik Wild, Susanne Yelin and Peter
Zoller for fruitful discussions.
\end{acknowledgments}

\appendix

\section{Classical Stability Analysis\label{app:classical}}

\subsection{Mathieu equation}

Performing a Taylor expansion for the electric field close to the
origin, $\mathrm{sin}\left(\tilde{x}\right)\approx\tilde{x}$, Eq.
(\ref{eq:newton}) can be mapped onto the well-known \textit{Mathieu}
differential equation by identifying the parameters appearing in the
standard \textit{Mathieu} differential equation,

\begin{equation}
\frac{d^{2}\tilde{x}}{d\tau^{2}}
+\left[a_{\mathrm{dc}}+2q\cos\left(2\tau\right)\right]\tilde{x}=0,
\label{eq:mathieu}
\end{equation}
as $a_{\mathrm{dc}}=0$ (no dc voltage) and
$q = V_\text{SAW}/E_{S}$.

In the case of vanishing dc contribution, according to Ref.\cite{paul90},
there is a stability zone for $0<q<q_{\mathrm{max}}$, with
$q_{\mathrm{max}}\approx0.92$, resulting in the maximum potential depth
of $V_{\mathrm{SAW}}=q_{\mathrm{max}}E_{S}$.
The lowest-order approximation to the ion trajectory $x\left(t\right)$
in the case $q^{2}\ll1$ is found to be
\begin{equation}
x\left(t\right)\approx2AC_{0}\underset{\mathrm{secular}}
{\underbrace{\cos\left(\beta\frac{\omega}{2}t\right)}}\underset{\mathrm{micromotion}}
{\underbrace{\left[1-\frac{q}{2}\cos\left(\omega t\right)\right]}},
\label{eq:lowest-order-classical-trajectory}
\end{equation}
where $\beta\approx q/\sqrt{2}$. If the fast low-amplitude oscillations
contained in in the second factor are neglected,
the secular motion can be approximated by that of a harmonic oscillator
with frequency $\omega_{0}=\beta\omega/2\ll\text{\ensuremath{\omega}}$.
The condition for the lowest-order approximation $q^{2}\ll1$ is
equivalent to a separation of timescales between secular and
micromotion, that is $\omega_{0}\ll\omega$. In this regime, the dynamics
can be described by an effective pseudopotential.

\subsection{Classical pseudopotential}

The classical dynamics in a high frequency field can be described
by an effective \textit{time-independent} Hamiltonian. Following
Refs.\cite{rahav03,rahav03b}, it can be calculated in a systematic
expansion in the inverse of the frequency $\omega$.
If the period of the force is small compared to the other time scales of
the problem, it is possible to separate the motion of the particle into
slow and fast parts.
This simplification is due to the fact that the particle does not have
sufficient time to react to the periodic force before this force changes
its sign.
Based on this separation of time scales, the motion for the slow part is
computed explicitly up to the order $\omega^{-4}$.
Note that the effective time-independent Hamiltonian depends on a
coordinate $X\left(t\right)$ which describes the slow part of the motion;
this coordinate is not the location of the particle, even though they are
almost identical at high frequencies $\omega$.
As outlined in Refs.\cite{rahav03,rahav03b}, the decomposition of $x(t)$
into slow and fast components can be written as
\begin{equation}
x\left(t\right)=X\left(t\right)+\xi\left(X,\dot{X},\omega t\right),
\end{equation}
where the fast part of the motion $\xi$ fulfills 
\begin{equation}
\bar{\xi}=\frac{1}{2\pi}\int_{0}^{2\pi}d\tau\xi\left(X,\dot{X},\tau\right)=0.
\end{equation}
By expanding $\xi$ in powers of $1/\omega$,
\begin{equation}
\xi=\sum_{i=1}^{\infty}\frac{1}{\omega_{i}}\xi_{i},
\end{equation}
such that Eq.(\ref{eq:mathieu}) leads to an equation for $X$ that is
time-independent and following Refs.\cite{rahav03,rahav03b}, we find the
following (classical) effective Hamiltonian describing the slow dynamics
$X(t)$
\begin{eqnarray}
H_{\mathrm{eff}}&=&\frac{P^{2}}{2m}\left[1+\frac{3}{8}q^{2}\cos^{2}\left(kX\right)\right] \nonumber \\
& &+ \frac{q}{8}V_{\mathrm{SAW}}\sin^{2}\left(kX\right)+\mathcal{O}\left(\omega^{-5}\right).\label{eq:Floquet-effective-classical-Hamiltonian}
\end{eqnarray}
Here, $P$ is the momentum conjugate to $X$. Given a solution $X\left(t\right)$,
the solution of the original problem can be obtained to appropriate
order of $1/\omega$ since $\xi$ is known explicitly in terms
of $X$ \cite{rahav03,rahav03b}. The pseudo-potential for the average
motion of the electron, $V_{\mathrm{eff}}=V_{0}\sin^{2}\left(kX\right)$,
with an amplitude given by 
\begin{equation}
V_{0}=\frac{q}{8}V_{\mathrm{SAW}}=\frac{q^{2}}{8}E_{S},\label{eq:classical-ponderomotive-potential} 
\end{equation}
is also referred to as ponderomotive potential \cite{leibfried03}.
Note that the correction to the kinetic term in
Eq.~\eqref{eq:Floquet-effective-classical-Hamiltonian} is a fourth-order
term, while the pseudo-potential $V_{\mathrm{eff}}$ is a second-order
contribution in $1/\omega$.
Close to the origin $x=0$, the effective potential $V_{\mathrm{eff}}$
can be approximated by a harmonic potential
$V_{\mathrm{eff}}\left(x\right)=(m/2)\omega_{0}^{2}x^{2}$ with an
oscillation frequency $\omega_{0}=\frac{q}{\sqrt{8}}\omega$, which is
equivalent to result obtained above from the Mathieu equation. 
Using this definition of the trapping frequency, the ponderomotive
potential becomes 
\begin{equation}
V_{\mathrm{eff}}=\left(\frac{\omega_{0}}{\omega}\right)^{2}E_{S}\sin^{2}\left(kX\right)
\end{equation}
We can then estimate the number of bound states $n_{\mathrm{b}}$ as 
\begin{equation}
n_{\mathrm{b}}\approx\frac{V_{0}}{\hbar\omega_{0}}=
\frac{1}{2}\sqrt{\frac{V_{0}}{E_{R}}},
\label{eq:number-of-bound-states-1}
\end{equation}
with the recoil energy $E_R = \hbar k^2 /2m$.

\subsection{Stability diagrams}

In this Appendix we provide further details on our classical stability
analysis. 
First, we would like to note that the stability diagrams shown in
Sec.\ref{sec:theory} are of approximate character as they were obtained
by interpolating our numerical results.
This is due to the deliberate choice of defining a stable trajectory in
terms of the maximal excursion during a sufficiently long propagation
time: two trajectories with almost equal parameters $q$ and
$k_{B}T/E_{S}$ can be judged as stable and unstable by this definition,
respectively, because only one of their amplitudes exceeds the cut-off
value set to one half of the lattice constant ($a/2$).
Second, the notion of (thermal) stability may be defined via the
mean-free path as well, by taking $l_\mathrm{mfp}$ as our cut-off value, 
in contrast to the trapping condition $\tilde x_{\text{max}} < \pi$. 
In that case, the regions of thermal stability increase as compared to
the ones shown in the main text, provided that $l_\mathrm{mfp}>a/2$.
The last inequality is likely to be fulfilled in high-mobility 2DEGs
where $l_\mathrm{mfp} \sim 10 \mu \mathrm{m}$.
Third, the stability analysis underlying Fig.~\ref{fig:stability}
neglects damping in the classical equation of motion; incorporating an
additional friction term may alter the notion of stability, since
particles which escape one lattice site can then be dynamically trapped
at a different lattice site.
Lastly, the state initialization via equipartition of thermal and
kinetic energies describes an average condition; in practice, only a
fraction of the electrons will fulfill this condition, where the details
depend on the statistical distribution of the initial conditions.
In order to estimate the statistical fraction of electrons whose
(initial) velocity $v$ is smaller than $v_0 = \sqrt{k_B T / m}$, given
by equipartition of thermal and kinetic energies of the particle, we
assume a Maxwell-Boltzmann distribution of velocities,
\begin{equation}
p(v) \mathrm{d} v = 2 \sqrt{ \frac{m}{2\pi k_B T} } \exp \left ( - \frac{mv^2}{2 k_B T} \right ) \mathrm{d} v,
\end{equation}
which yields $\int_0^{v_0} p(v) \mathrm{d} v \approx 0.68$; i.e., given
a thermal ensemble of particles we find that a significant fraction of
the electrons is found to be trapped.

\section{Quantum-Mechanical Floquet Analysis\label{app:qm-floquet}}

\textit{Preliminaries.}\textemdash We consider a quantum system with
a Hamiltonian that is periodic in time, $H\left(t+T\right)=H\left(t\right)$.
Floquet theory provides a natural framework to treat such a system
\cite{rahav03,rahav03b}.
The Bloch-Floquet theorem states that the eigenstates of the Schrödinger
equation
\begin{equation}
i\frac{\partial}{\partial t}\left|\Psi\right\rangle =H\left|\Psi\right\rangle ,
\end{equation}
obey the form 
\begin{equation}
\left|\Psi_{\lambda}\right\rangle =e^{-i\lambda t}\left|u_{\lambda}\left(\omega t\right)\right\rangle ,
\end{equation}
where $u_{\lambda}$ are periodic with respect to $\omega t$ with
period $2\pi$, that is
$u_{\lambda}\left(x,\omega\left(t+T\right)\right)=u_{\lambda}\left(x,\omega t\right)$
with $\omega=2\pi/T$.
The states $u_{\lambda}$ are called Floquet states and $\lambda$ is the
so-called quasienergy.
They have a natural separation into a slow part $e^{-i\lambda t}$ (with
the natural choice $0\leq\lambda<\omega$) and a fast part
$u_{\lambda}\left(x,\omega t\right)$.
Now, the goal is to find an effective description for the slow part of
the dynamics as was done above for the classical dynamics.
Formally, this is done by introducing a gauge transformation 
\begin{equation}
\left|\phi\right\rangle =e^{iF\left(t\right)}\left|\Psi\right\rangle ,
\end{equation}
where $F\left(t\right)$ is a Hermitian operator which is a periodic
function of time $t$, with the same period as $H\left(t\right)$,
such that the effective Hamiltonian $H_{\mathrm{eff}}$ in the Schrödinger
equation 
\begin{eqnarray}
i\frac{\partial}{\partial t}\left|\phi\right\rangle  & = & H_{\mathrm{eff}}\left|\phi\right\rangle ,\\
H_{\mathrm{eff}} & = & e^{iF}He^{-iF}+i\left(\frac{\partial}{\partial t}e^{iF}\right)e^{-iF},
\end{eqnarray}
is \textit{time-independent}. In particular, $H_{\mathrm{eff}}$ can
then be used to predict trapping due to oscillating potentials
\cite{rahav03}.

Typically, $F$ and $H_{\mathrm{eff}}$ cannot be computed exactly.
Following Refs.\cite{rahav03,rahav03b}, we expand $H_{\mathrm{eff}}$ and
$F$ in powers of $1/\omega$ and choose $F$ such that $H_{\mathrm{eff}}$
is time-independent to any given order.
In the following, we compute the effective Hamiltonian,
\begin{equation}
H_{\mathrm{eff}} = \sum_{n} \frac{1}{\omega^n} H_{\mathrm{eff}}^{(n)} ,\label{eq:heff-expansion}
\end{equation}
explicitly up to fourth order in $1/\omega$.

\subsection{Second order \label{subsec:Floquet-Theory:-Effective}}

Given the temporal periodicity of the driving only, it has been shown
\cite{rahav03,rahav03b} that the odd terms $H_{\mathrm{eff}}^{(1)}$,
$H_{\mathrm{eff}}^{(3)}$ from the perturbative expansion
\eqref{eq:heff-expansion} vanish.
Hence, the leading-order term (besides the purely kinetic contribution
$p^2/2m$) of the effective Hamiltonian is of second order in $1/ \omega$.

For the single-particle Hamiltonian under consideration,
\begin{equation}
H\left(t\right)=\frac{\hat{p}^{2}}{2m^{}}+V_{\mathrm{SAW}}\cos\left(\text{\ensuremath{\omega}t}\right)\cos\left(k\hat{x}\right),
\end{equation}
up to second order in $1/\omega$ we find
\begin{eqnarray}
H_{\mathrm{eff}} & = & \frac{\hat{p}^{2}}{2m^{}} + \frac{q}{8}V_{\mathrm{SAW}} \sin^{2}\left(k\hat{x}\right),\label{eq:Floquet-effective-quantum-Hamiltonian-2nd-order}
\end{eqnarray}
which is the second-order result given in Eq.~\eqref{eq:hamilt_eff}.
Hence, similar to the classical treatment, also within the quantum
mechanical Floquet framework, the effective potential, which is of
second order in the dimensionless coefficient $\omega_0/\omega$, can be
written as 
\begin{equation}
V_{\mathrm{eff}}\left(\hat{x}\right)=
\left(\frac{\omega_{0}}{\omega}\right)^{2}
E_{S}\sin^{2}\left(k\hat{x}\right).
\label{eq:effective-quantum-potential}
\end{equation}
Leading-order corrections to this result are of the order
$\mathcal{O}\left(\omega^{-4}\right)$.

\subsection{Fourth order \label{subsec:Floquet-Theory:fourth-order}}

Computing \eqref{eq:heff-expansion} explicitly up to
$\mathcal{O}(\omega^{-4})$ yields
\begin{eqnarray}
H_{\mathrm{eff}} & = & \frac{\hat{p}^{2}}{2m^{}} +
\frac{q}{8}V_{\mathrm{SAW}} \sin^{2}\left(k\hat{x}\right) \nonumber \\
& & + \frac{1}{2m^{}}\left[\hat{p}^{2}g\left(\hat{x}\right)+
2\hat{p}g\left(\hat{x}\right)\hat{p}+
g\left(\hat{x}\right)\hat{p}^{2}\right] \nonumber \\
& & +  \frac{q^{2}}{32}E_{R} \sin^{2}\left(k\hat{x}\right)+
\mathcal{O}\left(\omega^{-5}\right),
\label{eq:Floquet-effective-quantum-Hamiltonian-4th-order}
\end{eqnarray}
where 
\begin{equation}
g\left(\hat{x}\right)=\frac{3}{32}q^{2}\cos^{2}\left(k\hat{x}\right).
\label{eq:correction-term-kinetic-energy}
\end{equation}
In the classical limit (where $\hat{x}$ and $\hat{p}$ commute),
Eq.(\ref{eq:correction-term-kinetic-energy}) correctly reproduces the
kinetic correction term given in
Eq.(\ref{eq:Floquet-effective-classical-Hamiltonian}).
Compared to the classical result in
Eq.(\ref{eq:Floquet-effective-classical-Hamiltonian}),
Eq.\eqref{eq:Floquet-effective-quantum-Hamiltonian-4th-order} also
contains a fourth-order quantum-correction term which provides a
contribution to the pseudo-potential and which scales as
$\sim q^{2}E_{\mathrm{R}}$.
The eigenvalues of $H_{\mathrm{eff}}$ yield the Floquet quasienergies.
If the eigenstates of $H_{\mathrm{eff}}$ are known, then the Floquet
states can be computed up to order $\omega^{-4}$ using the explicit
expressions for $F$ derived in Refs.\cite{rahav03,rahav03b}.
Similarly to the classical analysis above, we find an effective
potential up to fourth order in $1/\omega$ which reads
\begin{eqnarray}
V_{\mathrm{eff}}\left(\hat{x}\right) & = & \left[\frac{q}{8}V_{\mathrm{SAW}}+\frac{q^{2}}{32}E_{R}\right]\sin^{2}\left(k\hat{x}\right),\\
 & = & \varepsilon^{2}E_{S}\sin^{2}\left(k\hat{x}\right),
\end{eqnarray}
where we have introduced the factor 
\begin{equation}
\varepsilon^{2}=\frac{q^{2}}{8}
\left[1+\tilde{q}\right],\,\,\,\,\,\,\,\,\,\,\tilde{q}=
\frac{E_{R}}{4E_{S}}=
\left(\frac{\hbar k}{2p_{s}}\right)^{2},
\label{eq:epsilon-perturbative-parameter}
\end{equation}
where the momentum $p_{s}$ is given by $p_{s}=m^{}v_{s}$. Within the
usual HO approximation, we obtain the corresponding trapping frequency
as 
\begin{equation}
\frac{\omega_{0}}{\omega}=\varepsilon=\frac{q}{2\sqrt{2}}\sqrt{1+\tilde{q}^{2}}.
\end{equation}

\section{Phonon-Induced Cooling in the Presence of Micromotion \label{app:cooling-micromotion}}

In this Appendix we discuss in detail the phonon-induced cooling-heating
dynamics and the resulting effective temperature of acoustically trapped
charge carriers, with full consideration of the time-dependence of
the SAW-induced trapping potential. Here, we focus on the relevant
decoherence processes due to coupling of the particle's motion to
the (thermal) phonon reservoir. Our analysis is built upon the master
equation formalism, a tool widely used in quantum optics for studying
the irreversible dynamics of a quantum system coupled to a macroscopic
environment. We detail the assumptions of our approach and discuss
in detail the relevant approximations.

\subsection{Time-Dependent System Dynamics\label{subsec:Time-Dependent-System-Dynamics}}

The system dynamics describing the motion of an electron (of mass $m$)
exposed to a SAW-induced standing wave is described by the Hamiltonian
given in  Eq.(\ref{eq:hamilt_sys}).
In the following, we restrict ourselves to the so-called Lamb-Dicke
regime in which the electron's motion is confined to a region much
smaller than the SAW wavelength $\lambda=2\pi/k$. The corresponding
approximation
$\cos\left(k\hat{x}\right)\approx\mathds1-\left(k^{2}/2\right)\hat{x}^{2}$
will be justified self-consistently below.
Dropping the first term $\sim\mathds1$ (which results in an irrelevant,
global phase only), the Hamiltonian $H_{S}\left(t\right)$ may be written
as 
\begin{equation}
H_{S}\left(t\right)\approx\frac{\hat{p}^{2}}{2m}+\frac{m}{2}W\left(t\right)\hat{x}^{2},\label{eq:system-Hamiltonian-approx}
\end{equation}
where $W\left(t\right)=-\left(\omega^{2}/2\right)q\cos\left(\omega t\right)$
can be identified as a time-varying spring constant, with the stability
parameter $q=V_{\mathrm{SAW}}/E_{S}$. In this form, the Hamiltonian
$H_{S}\left(t\right)$ and the corresponding dynamics have been studied
extensively in the literature (primarily in the context of trapped
ions), from both a classical and a quantum-mechanical point of view;
see for example Refs.\cite{cirac94,leibfried03,glauber92}. Still,
in order to set up the relevant notation for the subsequent analysis,
here we provide a self-contained discussion, closely following
Refs.\cite{cirac94,glauber92,leibfried03}.

Starting out from Eq.(\ref{eq:system-Hamiltonian-approx}), the Heisenberg
equations of motion for the electron's position $\hat{x}$ and momentum
operators $\hat{p}$ read 
\begin{eqnarray}
\dot{\hat{x}}\left(t\right) & = & \frac{1}{i\hbar}\left[\hat{x}\left(t\right),H_{S}\left(t\right)\right]=\hat{p}\left(t\right)/m,\\
\dot{\hat{p}}\left(t\right) & = & \frac{1}{i\hbar}\left[\hat{p}\left(t\right),H_{S}\left(t\right)\right]=-mW\left(t\right)\hat{x}\left(t\right),
\end{eqnarray}
which, when taken together, yield the well-known quantum Mathieu equation
\begin{equation}
\ddot{\hat{x}}\left(t\right)+W\left(t\right)\hat{x}\left(t\right)=0.
\end{equation}
This equation is equivalent to its classical counterpart if one replaces
the operator $\hat{x}\left(t\right)$ with a function $u\left(t\right)$
which satisfies the classical Mathieu equation
\cite{cirac94,glauber92,leibfried03}.
As well known in the context of trapped ions, stable solutions exist
only for certain values of the parameter $q$, which are usually defined
in terms of a stability chart; as compared to the standard analysis,
here we consider the simplified scenario without any dc voltage
\cite{cirac94}.
According to Floquet's theorem, such a stable solution $u\left(t\right)$
takes on the form 
\begin{equation}
u\left(t\right)=\sum_{n=-\infty}^{\infty}c_{2n}e^{i\left(\omega_{0}+n\omega\right)t}=e^{i\omega_{0}t}\Phi\left(t\right),
\end{equation}
where $\Phi\left(t\right)$ is a periodic function with period
$T=2\pi/\omega$, i.e. $\Phi\left(t+T\right)=\Phi\left(t\right)$.
Following Ref.\cite{leibfried03}, we consider solutions of the Mathieu
equation subject to the boundary conditions
\begin{equation}
u\left(0\right)=1,\,\,\,\,\,\dot{u}\left(0\right)=i\omega_{0}.
\end{equation}
As will be seen later, this choice of boundary conditions is convenient
for the appropriate definition of commutation relations. The (secular)
frequency $\omega_{0}/\omega$ is a function of $q$ and the coefficients
can be expressed in terms of a continued fraction; see e.g.
Refs.\cite{cirac94,leibfried03}.
In the limit $q^{2}\ll1$ it can be shown that $c_{0}\gg\left|c_{\pm2}\right|$,
such that the solution $u\left(t\right)$ is dominated by the so-called
secular frequency $\omega_{0}/\omega\approx q/\left(2\sqrt{2}\right)$,
which is much smaller than the driving frequency $\omega$. In the
corresponding pseudo-potential regime, a small-amplitude modulation
with micromotion frequency $\omega$ is superimposed on the slow (secular)
macro-motion.
To lowest order in $\sim q$, the solution $u\left(t\right)$ simplifies
to $u\left(t\right)=\exp\left[i\omega_{0}t\right]$, without accounting
for the micromotion.

Since the solution $u\left(t\right)$ and its complex conjugate
$u^{*}\left(t\right)$ form linearly independent solutions (which are
related to each other by the time-inversion symmetry inherent to the
Mathieu equation) \cite{kohler97,leibfried03}, they obey the Wronskian
identity
\begin{eqnarray}
\mathcal{W}\left(t\right) & = & u^{*}\left(t\right)\dot{u}\left(t\right)-u\left(t\right)\dot{u}^{*}\left(t\right),\\
 & = & u^{*}\left(0\right)\dot{u}\left(0\right)-u\left(0\right)\dot{u}^{*}\left(0\right),\\
 & = & 2i\omega_{0}.
\end{eqnarray}
The second equality simply follows from the fact that
$\mathcal{W}\left(t\right)$ is a constant of motion.
With this normalization, we obtain the sum rule 
\begin{equation}
\sum_{n}c_{n}^{2}\left(\frac{\omega_{0}+n\omega}{\omega_{0}}\right)=1.
\end{equation}
Since $\hat{x}\left(t\right)$ and $u\left(t\right)$ by definition
satisfy the same differential equation, one can construct an operator
$\hat{C}\left(t\right)$ which consists of an explicitly time-dependent
linear combination of the position and momentum operators as
\begin{equation}
\hat{C}\left(t\right)=i\sqrt{\frac{m}{2\hbar\omega_{0}}}\left[u\left(t\right)\dot{\hat{x}}\left(t\right)-\dot{u}\left(t\right)\hat{x}\left(t\right)\right],
\end{equation}
but which (being proportional to the Wronskian $\mathcal{W}$) turns
out to be a constant of motion \cite{cirac94,glauber92,leibfried03}.
Then, since 
\begin{equation}
\hat{C}\left(t\right)=\hat{C}\left(0\right)=\frac{1}{\sqrt{2m\hbar\omega_{0}}}\left[m\omega_{0}\hat{x}\left(0\right)+i\hat{p}\left(0\right)\right],
\end{equation}
one can readily identify $\hat{C}\left(t\right)$ with the well-known
annihilation operator associated with a \textit{static} harmonic oscillator
of mass $m$ and frequency $\omega_{0}$ as 
\begin{equation}
\hat{C}\left(t\right)=\hat{C}\left(0\right)=A,
\end{equation}
with the usual standard commutation relation 
\begin{equation}
\left[A,A^{\dagger}\right]=1.
\end{equation}
This static potential harmonic oscillator is usually referred to as
\textit{reference oscillator} \cite{leibfried03}. Since the operator
$A$ is time-independent, the same is true for
\begin{equation}
N=A^{\dagger}A,
\end{equation}
whose eigenstates are simply the familiar Fock states of the (static
potential) reference oscillator, with the standard ladder algebra
\begin{eqnarray}
A\left|n\right>_{\omega_{0}} & = & \sqrt{n}\left|n-1\right>_{\omega_{0}},\label{eq:annihilation-ref-oscillator}\\
A^{\dagger}\left|n\right>_{\omega_{0}} & = & \sqrt{n+1}\left|n+1\right>_{\omega_{0}},\label{eq:creation-ref-oscillator}
\end{eqnarray}
yielding directly $N\left|n\right>=n\left|n\right>_{\omega_{0}}$.

The Heisenberg operators $\hat{x}\left(t\right)$ and $\hat{p}\left(t\right)$
can then be expressed in terms of the classical Mathieu solutions
$u\left(t\right)$ as well as the (time-independent) creation and
annihilation operators of the reference oscillator as 
\begin{eqnarray}
\hat{x}\left(t\right) & = & \sqrt{\frac{\hbar}{2m\omega_{0}}}\left[u^{*}\left(t\right)A+u\left(t\right)A^{\dagger}\right],\\
\hat{p}\left(t\right) & = & \sqrt{\frac{\hbar m}{2\omega_{0}}}\left[\dot{u}^{*}\left(t\right)A+\dot{u}\left(t\right)A^{\dagger}\right].
\end{eqnarray}
Accordingly, the time dependence of the Heisenberg operators
$\hat{x}\left(t\right)$ and $\hat{p}\left(t\right)$ is captured entirely
by the classical Mathieu equation $u\left(t\right)$ and its complex
conjugate.
Note that
$\left[\hat{x}\left(t\right),\hat{p}\left(t\right)\right]=
\frac{\hbar}{2\omega_{0}}\mathcal{W}\left(t\right)=i\hbar$,
as desired.
For later reference, here we also define the Heisenberg operator for the
kinetic energy as 
\begin{eqnarray}
\frac{\hat{p}^{2}\left(t\right)}{2m} & = & \frac{\hbar}{4\omega_{0}}\left[\left|\dot{u}\left(t\right)\right|^{2}\left(A^{\dagger}A+AA^{\dagger}\right)\right.\nonumber \\
 &  & \left.+\left(\dot{u}^{*}\left(t\right)\right)^{2}A^{2}+\left(\dot{u}\left(t\right)\right)^{2}\left(A^{\dagger}\right)^{2}\right].\label{eq:Heisenberg-operator-kinetic-energy}
\end{eqnarray}

Since the annihilation (creation) operators $A$ $\left(A^{\dagger}\right)$
associated with the reference oscillator satisfy the usual algebra,
in complete analogy to the standard oscillator one may define a set
of basis states (in the Schrödinger picture) labeled as $\left|n;t\right>$
with $n=0,1,2,\dots$, which form the \textit{dynamic} counterpart
of the harmonic oscillator Fock states. The states $\left|n;t\right>$
are not stationary states, but do depend explicitly on time, as indicated
by the argument $t$ in the ket vector \cite{glauber92}. The ground
state of the reference oscillator $\left|n=0\right>_{\omega_{0}}$
obeys the condition 
\begin{equation}
A\left|n=0\right>_{\omega_{0}}=\hat{C}\left(t\right)\left|n=0\right>_{\omega_{0}}=0.
\label{eq:definition-ground-state-ref-oscillator}
\end{equation}
We can relate the Heisenberg operator $\hat{C}\left(t\right)$ to
its counterpart in the Schrödinger picture $\hat{C}_{S}\left(t\right)$
as $\hat{C}_{S}\left(t\right)=U\left(t\right)\hat{C}\left(t\right)U^{\dagger}\left(t\right)$,
with the unitary operator $U\left(t\right)$ which fulfills 
\begin{equation}
\dot{U}\left(t\right)=-iH_{S}\left(t\right)U\left(t\right).
\end{equation}
Explicitly, we find 
\begin{equation}
\hat{C}_{S}\left(t\right)=\frac{1}{2i}\left[\dot{u}\left(t\right)\sqrt{\frac{2m}{\hbar\omega_{0}}}\hat{x}-u\left(t\right)\sqrt{\frac{2}{m\hbar\omega_{0}}}\hat{p}\right].\label{eq:Floquet-shift-operator-explicit}
\end{equation}
Then, Eq.(\ref{eq:definition-ground-state-ref-oscillator}) can be
rewritten as 
\begin{equation}
\hat{C}_{S}\left(t\right)U\left(t\right)\left|n=0\right>_{\omega_{0}}=\hat{C}_{S}\left(t\right)\left|n=0;t\right>=0,
\end{equation}
where we have introduced the state $\left|n=0;t\right>$ in the
Schrödinger picture which evolves unitarily starting from the ground state of
the reference oscillator as
$\left|n=0;t\right>=U\left(t\right)\left|n=0\right>_{\omega_{0}}$.
The ladder-operator relations stated in
Eqs.(\ref{eq:annihilation-ref-oscillator}) and
(\ref{eq:creation-ref-oscillator}) for the reference oscillator can
easily be transferred to the Schrödinger picture, yielding 
\begin{eqnarray}
\hat{C}_{S}\left(t\right)\left|n;t\right> & = & \sqrt{n}\left|n-1;t\right>,\nonumber \\
\hat{C}_{S}^{\dagger}\left(t\right)\left|n;t\right> & = & \sqrt{n+1}\left|n+1;t\right>,\label{eq:Floquet-shift-operators}
\end{eqnarray}
implying $N_{S}\left(t\right)\left|n;t\right>=n\left|n;t\right>,$
with $N_{S}\left(t\right)=
\hat{C}_{S}^{\dagger}\left(t\right)\hat{C}_{S}\left(t\right)$.
Since the Schrödinger operators $\hat{C}_{S}\left(t\right)$,
$\hat{C}_{S}^{\dagger}\left(t\right)$ act as shift operators for the
Floquet states $\left|n;t\right>$, they will be referred to as Floquet
shift operators \cite{kohler97}.
Therefore, all other states forming the complete orthonormal basis
$\left\{ \left|n;t\right>\right\} $ can be constructed by repeated
operation on the ground state with the Schrödinger creation operator
$\hat{C}_{S}\left(t\right)$ (with the proper normalization) as 
\begin{equation}
\left|n;t\right>=\frac{[\hat{C}_{S}\left(t\right)]^{n}}{\sqrt{n!}}\left|n=0;t\right>.
\end{equation}
When expressing this equation in coordinate space, the micromotion
appears in the wavefunctions as a pulsation with the period $T=2\pi/\omega$
\cite{leibfried03}. Although the states $\left|n;t\right>$ are not
energy eigenstates (since they periodically exchange energy with the
driving field), they are typically referred to as \textit{quasi-stationary
states}, because for stroboscopic times (that are integer multiples
of the driving period $T$) the full evolution $U\left(t\right)$
boils down to multiplying the wavefunction by a simple phase factor
(as is the case for standard stationary states for all times). Because
of the periodicity of the micromotion, the quantum number $n$ (labeling
the quasi-energy states) can thus be tied to the electron's energy
averaged over a period $T=2\pi/\omega$ of the drive frequency. This
connection will be explored in greater detail below.

\subsection{The System-Bath Model}

While our previous discussion has exclusively focused on the time-dependent
system's dynamics {[}as described by the Hamiltonian $H_{S}\left(t\right)$
given in Eq.(\ref{eq:system-Hamiltonian-approx}){]}, in the following
we will develop a microscopic dissipative model, which describes the
electron's motional coupling to the (thermal) phonon reservoir.

The global Hamiltonian, describing both the electronic motion as well
as the phonon reservoir, can be formally decomposed as 
\begin{equation}
H\left(t\right)=H_{S}\left(t\right)+H_{B}+H_{I}.\label{eq:generic-system-bath-Hamiltonian}
\end{equation}
Here, the time-dependent system Hamiltonian $H_{S}\left(t\right)$
is given in Eq.(\ref{eq:system-Hamiltonian-approx}). The Hamiltonian
for the phonon bath $H_{B}$ is of the usual form 
\begin{equation}
H_{B}=\sum_{\mathbf{q},s}\omega_{\mathbf{q},s}a_{\mathbf{q},s}^{\dagger}a_{\mathbf{q},s},
\end{equation}
where $a_{\mathbf{q},s}^{\dagger}$ $\left(a_{\mathbf{q},s}\right)$
creates (annihilates) an acoustic phonon with wave vector
$\mathbf{q}=\left(\mathbf{q}_{||},q_{z}\right)$, polarization $s$ and
dispersion $\omega_{\mathbf{q},s}$.
Optical phonons can be disregarded at sufficiently low energies as
considered here \cite{golovach04}.
Following Ref.\cite{kornich14}, generically the electron-phonon
interaction takes on the form 
\begin{equation}
H_{I}=\sum_{\mathbf{q},s}W_{\mathbf{q},s}a_{\mathbf{q},s}e^{i\mathbf{q}\cdot\hat{\mathbf{r}}}+\mathrm{h.c.},
\end{equation}
with $\hat{\mathbf{r}}=\left(\hat{x},\hat{y},\hat{z}\right)$ denoting
the electron's three-dimensional position operator. The coupling constant
$W_{\mathbf{q},s}$ comprises both the deformation potential as well
as the piezoelectric coupling mechanism \cite{kornich14,golovach04};
it strongly depends on specific material properties, but can be left
unspecified for the sake of our discussion. For low-dimensional quasi-2D
systems as considered here, the Hamiltonian $H_{I}$ may be simplified
by projecting the electronic motional degrees of freedom onto the
lowest electronic orbital $\psi_{0}\left(z\right)$, leading to 
\begin{equation}
H_{I}\approx\sum_{\mathbf{q},s}\mathcal{F}\left(q_{z}\right)W_{\mathbf{q},s}a_{\mathbf{q},s}e^{i\mathbf{q}_{||}\cdot\hat{\mathbf{r}}_{||}}+\mathrm{h.c.},
\end{equation}
with the in-plane position operator
$\hat{\mathbf{r}}_{||}=\left(\hat{x},\hat{y}\right)$.
The form factor $\mathcal{F}\left(q_{z}\right)=
\int dze^{iq_{z}z}\left|\psi_{0}\left(z\right)\right|^{2}$
introduces a momentum cut-off, with $\mathcal{F}\left(q_{z}\right)$
approaching unity in the limit $\left|q_{z}\right|\ll d^{-1}$ and
vanishing for $\left|q_{z}\right|\gg d^{-1}$; here, $d\sim10\mathrm{nm}$
denotes the size of the quantum well along the $z$-axis \cite{golovach04}.
For the sake of clarity, here we will consider a quasi-one-dimensional
structure (a quantum wire) where the electron's motion is restricted
to the $x$-direction; compare our previous discussion in
Sec.\ref{subsec:Time-Dependent-System-Dynamics}.
In this case, the electron-phonon interaction reduces to 
\begin{equation}
H_{I}\approx\sum_{\mathbf{q},s}\tilde{W}_{\mathbf{q},s}a_{\mathbf{q},s}e^{iq\hat{x}}+\mathrm{h.c.},
\end{equation}
where the coupling $\tilde{W}_{\mathbf{q},s}$ accounts for transversal
confinement in both the $y$- and $z$-direction; moreover, we have
set $q=q_{x}$. Along the lines of
Sec.\ref{subsec:Time-Dependent-System-Dynamics}, again we restrict
ourselves to the Lamb-Dicke regime in which the electron's motion is
confined to a region much smaller than the wavelength of the relevant,
resonant phonon modes.
Then, taking $e^{iq\hat{x}}\approx\mathds1+iq\hat{x}$, and introducing
displaced bosonic bath modes as 
\begin{eqnarray}
b_{\mathbf{q},s} & = & -i\left(a_{\mathbf{q},s}+\tilde{W}_{\mathbf{q},s}^{*}/\omega_{\mathbf{q},s}\right),\\
b_{\mathbf{q},s}^{\dagger} & = & i\left(a_{\mathbf{q},s}^{\dagger}+\tilde{W}_{\mathbf{q},s}/\omega_{\mathbf{q},s}\right),
\end{eqnarray}
we finally arrive at the following microscopic system-bath model with
bilinear coupling between the system {[}as described by Eq.\ref{eq:system-Hamiltonian-approx}{]}
and a bath of noninteracting harmonic oscillators (i.e., the phonon
reservoir), 
\begin{eqnarray}
H_{B} & = & \sum_{\nu}\omega_{\nu}b_{\nu}^{\dagger}b_{\nu},\label{eq:bath-Hamiltonian-b}\\
H_{I} & = & \hat{x}\sum_{\nu}g_{\nu}\hat{x}_{\nu}+\hat{x}^{2}\sum_{\nu}\frac{g_{\nu}^{2}}{2m_{\nu}\omega_{\nu}^{2}}.\label{eq:bilinear-interaction-Hamiltonian}
\end{eqnarray}
where, to simplify the notation, we have introduced the multi-index
$\nu=\left(\mathbf{q},s\right)$ and $g_{\nu}$ specifies the coupling
strength between the system and each bath oscillator mode $\nu$.
Following the standard procedure in the literature, in
Eq.(\ref{eq:bilinear-interaction-Hamiltonian}) we have also included a
correction term which acts in the Hilbert space of the particle only and
compensates for a renormalization of the potential
$V\left(\hat{x},t\right)=\frac{m}{2}W\left(t\right)\hat{x}^{2}$
stemming from the system-reservoir coupling \cite{kohler97,breuer02}.
In this model, the reservoir spectral density, defined as 
\begin{equation}
J\left(\omega\right)=\pi\sum_{\nu}\frac{g_{\nu}^{2}}{2m_{\nu}\omega_{\nu}}\delta\left(\omega-\omega_{\nu}\right),
\end{equation}
encodes all features of the environment relevant for the reduced system
description \cite{kohler97}.

\subsection{Quantum Master Equation, Quasi-Stationary State and Effective Temperature}

The time-dependent, dissipative quantum system described by
Eqs.(\ref{eq:generic-system-bath-Hamiltonian}),
(\ref{eq:system-Hamiltonian-approx}), (\ref{eq:bath-Hamiltonian-b})
and (\ref{eq:bilinear-interaction-Hamiltonian}), commonly referred
to as parametrically driven, dissipative harmonic quantum oscillator,
has been studied in great detail previously in Ref.\cite{kohler97}.
Within one unified Born-Markov and Floquet framework, the authors
of Ref.\cite{kohler97} have derived a quantum Master equation for
the electronic motion, fully taking into account the explicit
time-dependence of the system Hamiltonian $H_{S}\left(t\right)$.

\textit{Master equation.}\textemdash By tracing out the unobserved
degrees of freedom of the phonon reservoir Kohler \textit{et al.}
derive an effective equation of motion for the reduced, electronic
density matrix $\rho$, which is irreversibly coupled to a thermal
phonon reservoir \cite{kohler97}. In addition to the standard assumptions
of a weak system-reservoir coupling (Born approximation), and a short
reservoir correlation time (Markov approximation), the analysis has
been restricted to an ohmic spectral density where
$J\left(\omega\right)\sim\omega$ (which, however, may be generalized to
a more general setting straight-forwardly).
Under these conditions, the central master equation can be written as 
\begin{eqnarray}
\dot{\rho} & = & -\frac{i}{\hbar}\left[\hat{H}_{S}\left(t\right),\rho\right]+\mathcal{L}_{\gamma}\rho,\label{eq:Kohler-QME-Schroedinger-picture}
\end{eqnarray}
with 
\begin{equation}
\mathcal{L}_{\gamma}\rho=\gamma\left(N+1\right)\mathscr{D}\left[\hat{C}_{S}\left(t\right)\right]\rho+\gamma N\mathscr{D}\left[\hat{C}_{S}^{\dagger}\left(t\right)\right]\rho.
\end{equation}
Here, $\mathscr{D}\left[c\right]\rho=c\rho c^{\dagger}-\frac{1}{2}\left\{ c^{\dagger}c,\rho\right\} $
is a dissipator of Lindblad form, $\gamma$ denotes the effective,
incoherent damping rate due to coupling to the thermal phonon reservoir,
and
\begin{equation}
N=\sum_{n}c_{2n}^{2}\frac{\omega_{0}+n\omega}{\omega_{0}}\bar{n}_{\mathrm{th}}\left(\omega_{0}+n\omega\right),
\end{equation}
with $\bar{n}_{\mathrm{th}}\left(\omega\right)=\left(\exp\left[\hbar\omega/k_{B}T\right]-1\right)^{-1}$,
refers to a generalized effective thermal-bath occupation number.
Note that Eq.(\ref{eq:Kohler-QME-Schroedinger-picture}) retains the
periodicity of the driving and exhibits Lindblad form. Moreover, the
dissipative part of Eq.(\ref{eq:Kohler-QME-Schroedinger-picture})
is of the same form as for the well-known undriven dissipative harmonic
oscillator, with the Floquet shift operators defined in
Eqs.(\ref{eq:Floquet-shift-operator-explicit}) and
(\ref{eq:Floquet-shift-operators}) replacing the usual creation
and annihilation operators. Note that in the pseudopotential limit
(where $c_{0}$ is much larger than all other Floquet coefficients)
the effective thermal occupation reduces to
$N=\bar{n}_{\mathrm{th}}\left(\omega_{0}\right)$, that is the standard
bosonic thermal occupation at the secular frequency $\omega_{0}$.

The Master equation given in Eq.(\ref{eq:Kohler-QME-Schroedinger-picture})
is valid provided that the following conditions are satisfied \cite{kohler97}:
(i) First, the Markov approximation is satisfied provided that autocorrelations
of the bath (which typically decay on a timescale $\sim\hbar/k_{B}T$)
decay quasi instantaneously on the timescale of system correlations
$\sim\gamma^{-1}$. In principle, the damping rate $\gamma$ should
be replaced by the thermally enhanced rate $\gamma_{\mathrm{eff}}=\gamma\left(N+1\right)$;
however, we will be interested mostly in the low-temperature, pseudopotential
regime where $\gamma_{\mathrm{eff}}\approx\gamma$. Thus, the Markov
approximation yields the condition $\hbar\gamma\ll k_{B}T$. (ii)
Second, the (weak-coupling) Born approximation holds provided that
the dissipative damping rate $\gamma$ is small compared to the relevant
system's transition frequencies, yielding the requirement $\gamma\ll\omega_{0}$.
Taking together conditions (i) and (ii) (and setting $\hbar=1$ for
the moment) gives the requirement 
\begin{equation}
\gamma\ll\omega_{0},k_{B}T,
\end{equation}
which (as shown below) comprises the regime for ground-state cooling
where $\gamma\ll k_{B}T\ll\omega_{0}$. (iii) Finally, when deriving
Eq.(\ref{eq:Kohler-QME-Schroedinger-picture}), the reservoir spectral
density $J\left(\omega\right)$ has been assumed to be ohmic (i.e.,
$J\left(\omega\right)\sim\omega$).


\textit{Quasi-stationary state.}\textemdash Using Eq.(\ref{eq:Floquet-shift-operators}),
the (asymptotic) quasi-stationary solution $\rho_{\mathrm{ss}}\left(t\right)$
associated with the Master equation (\ref{eq:Kohler-QME-Schroedinger-picture})
is readily found to be 
\begin{equation}
\rho_{\mathrm{ss}}\left(t\right)=\frac{1}{N+1}\sum_{n=0}^{\infty}\left(\frac{N}{N+1}\right)^{n}\left|n;t\right>\left<n;t\right|,
\end{equation}
where $\left|n;t\right>$ refer to the generalized (time-dependent)
Fock states as discussed above \cite{kohler97}. The quasi-stationary
solution $\rho_{\mathrm{ss}}\left(t\right)$ is dark with respect
to the phonon-induced dissipation, that is
$\mathcal{L}_{\gamma}\rho_{\mathrm{ss}}\left(t\right)=0$ for all times,
and, being a mixture of the Floquet solutions $\left|n;t\right>$,
evolves periodically with the period of the driving field, i.e.,
$\rho_{\mathrm{ss}}\left(t+T\right)=\rho_{\mathrm{ss}}\left(t\right)$.

While the notion of temperature becomes ambiguous for an explicitly
time-dependent problem as considered here, in the following we will
adopt the reasoning presented in Ref.\cite{cirac94} and take the
mean kinetic energy (defined as the quantum kinetic energy
$\overline{\left\langle \hat{p}^{2}(t)\right\rangle }/2m$, time-averaged
over one period $T=2\pi/\omega$ of the fast micromotion) as our figure of
merit for assessing the cooling-heating dynamics in more detail.
To do so, let us first transform our analysis into a frame that is moving
with the electron. Formally, this transformation is defined as
$\varrho=U^{\dagger}\left(t\right)\rho U\left(t\right)$, with the unitary
operator that satisfies 
\begin{equation}
\dot{U}\left(t\right)=-iH_{S}\left(t\right)U\left(t\right).
\end{equation}
Then, in the corresponding interaction picture (which coincides with the
Heisenberg picture defined in Sec.\ref{subsec:Time-Dependent-System-Dynamics})
the dynamics described by Eq.(\ref{eq:Kohler-QME-Schroedinger-picture})
reduces to a purely dissipative master equation,
$\dot{\varrho}=\mathcal{L}\varrho$,
\begin{equation}
\dot{\varrho}=\gamma\left(N+1\right)\mathscr{D}\left[A\right]\varrho+\gamma N\mathscr{D}\left[A^{\dagger}\right]\varrho,
\end{equation}
where $A$ $\left(A^{\dagger}\right)$ refers to the time-independent
annihilation (creation) operator associated with the reference oscillator
discussed in Sec.\ref{subsec:Time-Dependent-System-Dynamics}. Since
the Liouvillian $\mathcal{L}$ is time-independent, one can easily
explain the phonon-induced cooling dynamics via the eigenstates of
$A^{\dagger}A$, as defined in Sec.\ref{subsec:Time-Dependent-System-Dynamics}.
For simplicity, let us focus on the pseudopotential regime where
$N\approx\bar{n}_{\mathrm{th}}\left(\omega_{0}\right)$ as discussed above;
then, for sufficiently low temperatures $\left(k_{B}T\ll\hbar\omega_{0}\right)$
the cooling dynamics dominate over the heating processes such that,
at the end of the cooling process, we have
$\left\langle A^{\dagger}A\right\rangle =\left\langle A^{2}\right\rangle
=\left\langle A^{\dagger}A^{\dagger}\right\rangle =0$.
In this regime the expectation value for the quantum kinetic energy
reduces to 
\begin{equation}
\frac{\left\langle \hat{p}^{2}\left(t\right)\right\rangle }{2m}\overset{\mathrm{cooling}}{\longrightarrow}\frac{\hbar}{4\omega_{0}}\left|\dot{u}\left(t\right)\right|^{2},
\end{equation}
as one can readily deduce from Eq.(\ref{eq:Heisenberg-operator-kinetic-energy}).
Averaging this expression (which still fully accounts for the time
dependence of the potential) over one micromotion period, we obtain
\begin{eqnarray}
\frac{\overline{\left\langle \hat{p}^{2}(t)\right\rangle }}{2m} & = & \frac{\hbar}{4\omega_{0}}\sum_{n}\left|c_{2n}\right|^{2}\left(\omega_{0}+n\omega\right)^{2},\label{eq:averaged-kinetic-energy}\\
 & = & \frac{\hbar\omega_{0}}{4}+\Delta_{\mathrm{heat}},
\end{eqnarray}
where we have separated the residual kinetic zero-point motion in
the ground state of the secular reference oscillator
$\sim\hbar\omega_{0}/4$ from the non-zero heating term 
\begin{equation}
\Delta_{\mathrm{heat}}=\frac{\hbar}{4\omega_{0}}\sum_{n\neq0}\left|c_{2n}\right|^{2}\left(\omega_{0}+n\omega\right)^{2},
\end{equation}
which may be viewed as micromotion-induced heating. While the full
expression given in Eq.(\ref{eq:averaged-kinetic-energy}) can be
evaluated numerically using the well-known solutions of the Mathieu
equation (compare for example Ref.\cite{cirac94}), a simple estimate
(for $q^{2}\ll1$) shows $\Delta_{\mathrm{heat}}\gtrsim\hbar\omega_{0}/4$.
Therefore, in agreement with the results presented in Ref.\cite{cirac94}
for trapped ions, we then find
$\overline{\left\langle \hat{p}^{2}(t)\right\rangle }/2m\gtrsim\hbar\omega_{0}/2$
for the time-averaged kinetic energy in the pseudo-potential regime,
which coincides with twice the residual kinetic zero-point motion
in the ground state of the reference oscillator.

In conclusion, our analysis shows that micromotion does lead to some
heating as compared to the naive estimate based on the slow secular
motion only, but (in the pseudopotential regime of interest, where
$q^{2}\ll1$) this apparent heating mechanism is strongly suppressed and
amounts to merely a factor of 2 increase only in the particle's time-averaged
kinetic energy.

\subsection{Exact Numerical Simulations and Discussion}

\textit{Setup.}\textemdash Since the electronic dynamics described
by Eq.(\ref{eq:Kohler-QME-Schroedinger-picture}) are purely Gaussian,
an exact solution is feasible. Therefore, in the following we will
complement our analytical findings with numerically exact simulations
for the electron's dynamics. Based on Eq.(\ref{eq:Kohler-QME-Schroedinger-picture}),
one can readily derive a closed dynamical equation 
\begin{equation}
\frac{d}{dt}\mathbf{v}=\mathcal{M}\left(t\right)\mathbf{v}+\mathbf{C}\left(t\right),
\end{equation}
where $\mathbf{v}$ is a five-component vector comprising the first-
and second-order moments, that is
$\mathbf{v}=\left(\left\langle \hat{x}\right\rangle _{t},
\left\langle \hat{p}\right\rangle _{t},
\left\langle \hat{x}^{2}\right\rangle _{t},
\left\langle \hat{p}^{2}\right\rangle _{t},
\left\langle \hat{x}\hat{p}+\hat{p}\hat{x}\right\rangle _{t}\right)^{\top}$.
Since the first- and second-order moments are decoupled, the dynamical
matrix $\mathcal{M}$ is of block-diagonal form.

\textit{Numerical results.}\textemdash
As illustrated in Fig.\ref{fig:simulation-gaussian-dynamics},
in the regime $q^{2}\ll1$ we numerically find that (i) the electronic
motion can be described very well by a simple damped harmonic oscillator
with secular frequency $\omega_{0}$, (ii) the electronic motion is
cooled by the phonon reservoir and (iii) the Lamb-Dicke approximation
is well satisfied. Let us elaborate on these statements in some more
detail: (i) When disregarding micromotion, the dynamics can approximately
be described by a simple damped harmonic oscillator with secular frequency
$\omega_{0}$. As shown in Fig.\ref{fig:simulation-gaussian-dynamics},
the effective, time-independent master equation 
\begin{eqnarray}
\dot{\rho} & = & -i\omega_{0}\left[a^{\dagger}a,\rho\right]+\gamma\left(\bar{n}_{\mathrm{th}}\left(\omega_{0}\right)+1\right)\mathscr{D}\left[a\right]\rho\nonumber \\
 &  & +\gamma\bar{n}_{\mathrm{th}}\left(\omega_{0}\right)\mathscr{D}\left[a^{\dagger}\right]\rho,
\end{eqnarray}
with $a$ $\left(a^{\dagger}\right)$ denoting the usual annihilation
(creation) operators for the canonical harmonic oscillator, captures
well the most pertinent features of the electronic dynamics, provided
that $q^{2}\ll1$; compare the dashed orange line in
Fig.\ref{fig:simulation-gaussian-dynamics}.
(ii) As suggested by our analytical analysis, the phonon reservoir
provides an efficient cooling mechanism for the electron provided
that the host temperature is sufficiently low, that is
$k_{B}T\ll\omega_{0}$.
(iii) Regarding the last statement (iii) we have numerically verified
that both the expectation value for the electron's motion as well
as the corresponding fluctuations are small compared to the SAW
wavelength $\lambda=2\pi/k$, i.e.,
$k\left\langle \hat{x}\right\rangle _{t}\ll1$ and
$k\sigma_{x}\ll1$ with $\sigma_{x}^{2}=
\left\langle \hat{x}^{2}\right\rangle _{t}-\left\langle \hat{x}\right\rangle _{t}^{2}$.
Furthermore, the Lamb-Dicke approximation underlying the bilinear
system-bath interaction Hamiltonian {[}compare
Eq.(\ref{eq:bilinear-interaction-Hamiltonian}){]} can be justified as
follows: Since the effective transition frequency
$\omega_{0}$ is much smaller than the SAW driving frequency
($\omega_{0}=\varepsilon\omega$, with $\varepsilon\ll1$), the same is
true for the relevant phonon wavenumber $k_{0}$.
Using the relation $\omega_{0}=v_{s}^{(b)}k_{0}$ (where $v_{s}^{(b)}$
refers to the speed of sound associated with some relevant bulk phonon
mode), the latter can be expressed as
$k_{0}=\varepsilon(v_{s}/v_{s}^{(b)})k$ (with $v_{s}$ denoting the speed
of sound of the SAW mode driven by the IDTs, as usual).
Therefore, even for higher Rayleigh SAW-modes whose speed of sound
$v_{s}$ may exceed the lowest value of $v_{s}^{(b)}$, our approximate
treatment of the system-bath Hamiltonian is well justified, provided
that $v_{s}\lesssim v_{s}^{(b)}/\varepsilon$ holds.
Note that material-engineering strategies as discussed in the main text
would increase $v_{s}$ in the same way as $v_{s}^{(b)}$, providing a
very good justification for our linearized Hamiltonian
(\ref{eq:bilinear-interaction-Hamiltonian}) since
$k_{0}\lesssim\varepsilon k$.

Finally, the parameter regime of interest is summarized and discussed
extensively in Section  \ref{sec:theory} of the main text, while the
experimental feasibility thereof is discussed in Section
\ref{sec:implementation}. 

\section{Case Study \& Practical Considerations}  \label{sec:case-study}

In this Appendix we provide further details regarding several practical considerations 
that are relevant for a faithful experimental realization of our proposal.
First, we provide comprehensive overview of the key quantities of our analysis and self-consistent estimates thereof. 
Next, we address microwave-induced heating effects.
Lastly, we discuss electron spin decoherence effects due to (nuclear) spin noise. 

\begin{table*}
\begin{tabular}{|c|c|c|c|c|c|c|c|c|c|}
\hline 
$\hbar \omega [\mu \mathrm{eV}]$ & $q=V_\mathrm{SAW}/E_S$ & $\hbar \omega_0 [\mu \mathrm{eV}]$ & $V_0 [\mu \mathrm{eV}]$ & $n_{b} = V_0/\hbar \omega_0$ & $a = \lambda/2 [\mathrm{nm}]$ & $d [\mathrm{nm}]$ & $t \ [\mu \mathrm{eV}]$ & $U [\mu \mathrm{eV}]$ & $k_B T [\mu \mathrm{eV}]$ \tabularnewline
\hline 
\hline 
207 & 0.5 - 0.7 & 37-51  & 31-61 & 0.85-1.2 & 180 & 10-100 & 0.7-1.8 & 5-270 & 1-10\tabularnewline
\hline 
\end{tabular}
\caption{\label{table:case-study}Important (energy) scales for an exemplary setup with $E_S = 1 \mathrm{meV}$
and $f = 50 \mathrm{GHz}$. $d$ denotes the distance between the screening layer and the 2DEG.}
\end{table*}

\textit{Case study}.\textemdash Typical parameter regimes for the key quantities of our analysis are
given in Table~\ref{table:case-study}.
The parameters are chosen self-consistently with respect to the
requirements derived in the main text, see
Eq.~\ref{eq:parameter-regime-of-interest-1}.
Note that the high-SAW frequencies lead to large energy scales in the
effective (harmonic-oscillator) problem.
For comparison, ions are typically confined in traps with
harmonic-oscillator energy $\hbar \omega_{0}\sim10\mathrm{MHz}$
\cite{leibfried03}. 
For the SAW velocity $v_s$, we assume an ultra-fast PSAW mode in
AlN/diamond ($v_s \approx 18 \mathrm{km/s}$) as described in the main
text and a corresponding effective hole mass $m = 1.1 m_0$ in the host
material GaN where the 2DEG is located.


\textit{Heating}.\textemdash
In order to avoid excessive heating 
of the effective electron temperature above the dilution fridge temperature
in the presence of RF driving, we (i) either need the heat dissipation
$W_{\mathrm{heat}}$ to be balanced by the applied cooling power $P_{\mathrm{cool}}$
(for which, in an actual experiment, the way the sample is heat sunk
is very important) and/or (ii) the heat dissipation to be too slow
to change the electron's temperature on relevant experimental timescales
after the IDT induced driving has been turned on. In the following
we argue why the requirements (i) and (ii) can both be fulfilled under
realistic conditions: (i) First, recall that our proposal is based
on low power SAWs (as a direct consequence of the limitations imposed
by Mathieu's equation) \cite{footnote-amplitude}. 
Since
the potential amplitude due to a single IDT is limited by Mathieu-like
stability arguments as $V_{\mathrm{IDT}}=V_{\mathrm{SAW}}/2=(q/2)E_{S}\lesssim0.5\mathrm{meV}$,
the proposed setup operates at SAW-induced amplitudes that are about
two orders of magnitude smaller than what is common for SAW-induced
electron transport experiments (where typically $V_{\mathrm{IDT}}\approx40\mathrm{meV}$
\cite{furuta04,hermelin11}). Based on experimental results presented
in Refs.\cite{schnebele06,hermelin11,utko06}, we find that SAW amplitudes
$V_{\mathrm{IDT}}\approx1\mathrm{meV}$ can be reached with an applied
RF power $P\approx-10\mathrm{dBm}(0.1\mathrm{mW})$, in the desired
SAW frequency range $\omega/2\pi\approx30\mathrm{GHz}$ (as needed
to enter the pseudo-potential regime), whereas high-amplitude electron
transport measurements operate at $P>+10\mathrm{dBm}(10\mathrm{mW})$
\cite{hermelin11}. This estimate is based on experiments with relatively
wide IDTs in GaAs; therefore, the power budget $P$ could be further
reduced (if needed) by reducing the width $W$ of the IDTs (which
is typically several hundreds of $\mu\mathrm{m}$ long \cite{manenti16},
i.e., much longer than necessarily required for an acoustic trap or
lattice) and/or using strongly piezoelectric materials \cite{morgan07,wu14,datta86}
where the electro-mechanical coupling efficiency is much larger than
for the weakly piezoelectric material GaAs. Heating effects as a function
of the applied RF power $P$ have been investigated experimentally
in detail in Refs.\cite{utko06,schnebele06}: Here, at a comparatively
large microwave power $P=+5\mathrm{dBm}$ the SAW-induced heating
has been measured to be $W_{\mathrm{heat}}^{\mathrm{SAW}}\approx0.1\mathrm{mW}$.
We may estimate this source of heating as $W_{\mathrm{heat}}^{\mathrm{SAW}}\approx\hbar\omega\times\left(V_{\mathrm{SAW}}/V_{0}\right)^{2}\kappa$,
where $\hbar\omega$ is the energy of a single phonon and the second
factor gives the total phonon loss rate in terms of the phonon number
$N_{\mathrm{ph}}\approx\left(V_{\mathrm{SAW}}/V_{0}\right)^{2}$ and
the decay rate $\kappa=\omega/Q$; here, $V_{0}$ refers to the amplitude
associated with a single phonon \cite{schuetz15} and $Q$ is the
quality factor associated with the driven SAW mode. 
However, it has been shown in Ref.\cite{utko06} that
$W_{\mathrm{heat}}^{\mathrm{SAW}}$ accounts for $\sim10\%$ of the
overall heating only, due to the limited efficiency of the IDTs. While
this ratio may be improved with more sophisticated IDT designs \cite{ekstroem17,datta86,morgan07},
an overall heating of $W_{\mathrm{heat}}\approx10W_{\mathrm{heat}}^{\mathrm{SAW}}\approx1\mathrm{mW}$
is still compatible with the cooling power of state-of-the-art dilution
refrigerators, which can reach $P_{\mathrm{cool}}=1\mathrm{mW}$ at
$T\approx100\mathrm{mK}$ \cite{vandersypen16}; here, to maximize
the cooling efficiency in an actual experiment, attention should be
paid to the the specific way the sample is heat sunk. Since the proposed
AL setup operates at much lower RF power levels {[}$P\lesssim-10\mathrm{dBm}(0.1\mathrm{mW})$
as compared to $P=+5\mathrm{dBm}(3\mathrm{mW})${]}, the overall heat
dissipation $W_{\mathrm{heat}}$ can be balanced by the applied
cooling power $P_{\mathrm{cool}}$ for the specific parameters under
consideration. This finding is further supported by the experiments
presented in Refs.\cite{utko06,schnebele06}, where for low-power
SAWs no significant heating above the base temperature has been observed.
(ii) Second, the IDTs generating the SAWs can be placed very far away
from the center of the trap, without losing acoustic power, thereby
reducing local heat dissipation near the center of the trap due to
the applied RF power. For example, in Ref.\cite{hermelin11} (and
many similar setups) the SAW transducer has been placed approximately
2mm away from the center of the sample. In this way, the dominant
local heating at the IDT may be suppressed efficiently, at least on
timescales that are short compared to the one set by the material-specific
thermal diffusivity (which specifies the rate of transfer of heat
from the IDT to the cold center of the trap). While this timescale
is strongly material-dependent, a rough estimate for GaAs shows that
it can lie in the millisecond range (for IDTs placed $\sim1\mathrm{mm}$
away from the center of the trap), which is much longer than any relevant
experimental timescale. This reasoning is also in line with experimental
results showing that the effective temperature increase could be further
reduced when using pulsed schemes rather than CW \cite{schnebele06};
note that this approach is fully compatible with our discussion on optimized driving
schemes. 
In summary, we conclude that for realistic cooling powers
and/or IDTs placed sufficiently far away from the center of the trap
microwave induced heating effects should not lead to a significant
increase of the effective particle temperature (as compared to the
base temperature) since the AL setup is based on low amplitude SAWs
with $V_{\mathrm{IDT}} \lesssim 0.5\mathrm{meV}$, as a direct consequence
of the Mathieu-type stability arguments. 

\textit{Nuclear spin noise}.\textemdash The observation of
coherent spin physics as outlined in Sec.\ref{sec:Many-Body-Physics}
may be impeded by electron spin decoherence.
For GaAs-based systems, the electron spin coherence timescale will be
largely limited by the relatively strong hyperfine interaction between
the electronic spin and the nuclei in the host environment
\cite{chekhovich13}, resulting in a random, slowly evolving magnetic
(Overhauser) field for the electronic spin, and eventually leading to a
loss of spin coherence on a timescale $\sim T_{2}^{\star}$.
The latter depends on the number of nuclear spins the electron
effectively interacts with.
Since the electron's spatial extension
$\Delta x/a\approx1/\left(\pi\sqrt{2n_{b}}\right)$ is comparable to the
typical size of gate-defined quantum dots for realistic parameter values,
we estimate $T_{2}^{\star}\sim15\mathrm{ns}$ \cite{chekhovich13}.
Then, in the first approximation, the detrimental effects due to
Overhauser noise may be neglected provided that the condition
$J\gg1/T_{2}^{\star}$ is fulfilled, i.e., if coherent spin exchange
$\sim1/J$ is much faster than electron spin dephasing.
According to our estimates provided above this regime is within reach
even for GaAs-based systems, where electron spin dephasing is known to
be relatively fast \cite{chekhovich13}.
In this respect, even more promising estimates apply to nuclear spin
free systems such as $^{28}\mathrm{Si/SiGe}$ where the influence of
nuclear spins on the electron spins is largely eliminated
\cite{zwanenburg13}.
While such a silicon-based setup will require a more sophisticated
heterostructure including some piezoelectric layer on top (as has been
studied experimentally in Ref.\cite{buy12}), it should profit from
significantly prolonged dephasing times $T_{2}^{\star}>100\mu\mathrm{s}$
\cite{veldhorst14}.
Finally, as argued for example in Ref.\cite{mcNeill11}, Overhauser-field induced
spin dephasing can be suppressed based on motional-narrowing techniques, when
moving around the acoustic dots (lattice sites) such that the electron effectively samples many 
different Overhauser fields. 




\begin{thebibliography}{10}

\bibitem{ashkin00}A. Ashkin, \textit{History of optical trapping and manipulation of small-neutral particle, atoms, and molecules},
IEEE J. Sel. Top. Quantum Electron. \textbf{6}, 841 (2000). 

\bibitem{lang03}M. J. Lang and S. M. Block,\textit{Laser-based optical tweezers},
Am. J. Phys. \textbf{71}, 201 (2003). 

\bibitem{anderson95}M. H. Anderson, J. R. Ensher, M. R. Matthews, C. E.
Wieman, and E. A. Cornell, \textit{Observation of bose-einstein condensation in a dilute atomic vapor},
Science \textbf{269}, 198 (1995).

\bibitem{bradley95}C. C. Bradley, C. A. Sackett, J. J. Tollett, and R. G. Hulet,
\textit{Evidence of Bose-Einstein Condensation in an Atomic Gas with Attractive Interactions},
Phys. Rev. Lett. \textbf{75}, 1687 (1995).

\bibitem{davis95}K. B. Davis, M.-O. Mewes, M. R. Andrews, N. J.
van Druten, D. S. Durfee, D. M. Kurn, and W. Ketterle,
\textit{Bose-Einstein Condensation in a Gas of Sodium Atoms},
Phys. Rev. Lett. \textbf{75}, 3969 (1995).

\bibitem{blatt12}R. Blatt and C. F. Roos,
\textit{Quantum simulations with trapped ions},
Nat. Phys. \textbf{8}, 277 (2012).

\bibitem{bloch08}I. Bloch, J. Dalibard, W. Zwerger,
\textit{Many-body physics with ultracold gases},
Rev. Mod. Phys. \textbf{80}, 885 (2008).

\bibitem{bloch12}I. Bloch, J. Dalibard, and S. Nascimbène,
\textit{Quantum simulations with ultracold quantum gases},
Nat. Phys. \textbf{8}, 267 (2012).

\bibitem{hanson07}R. Hanson, L. P. Kouwenhoven, J. R. Petta, S. Tarucha, and
L. M. K. Vandersypen,
\textit{Spins in few-electron quantum dots},
Rev. Mod. Phys. \textbf{79}, 1217 (2007).

\bibitem{schuetz10}M. J. A. Schuetz, M. G. Moore and C. Piermarocchi,
\textit{Trionic optical potential for electrons in semiconductors},
Nature Phys. \textbf{6}, 919 (2010).

\bibitem{cunningham99}J. Cunningham, V. I. Talyanskii, J. M. Shilton, M. Pepper, M. Y. Simmons, and D. A. Ritchie,
\textit{Single-electron acoustic charge transport by two counterpropagating surface acoustic wave beams},
Phys. Rev. B \textbf{60}, 4850 (1999).

\bibitem{stotz05}J. A. H. Stotz, R. Hey, P. V. Santos, and K. H. Ploog,
\textit{Coherent spin transport through dynamic quantum dots},
Nature Mat. \textbf{4}, 585 (2005).

\bibitem{hermelin11}S. Hermelin, S. Takada, M. Yamamoto, S. Tarucha,
A. D. Wieck, L. Saminadayar, C. Bäuerle, and T. Meunier,
\textit{Electrons surfing on a sound wave as a platform for quantum optics with flying electrons},
Nature \textbf{477}, 435 (2011).

\bibitem{mcNeill11}R. P. G. McNeil, M. Kataoka, C. J. B. Ford, C.
H. W. Barnes, D. Anderson, G. A. C. Jones, I. Farrer, and D. A. Ritchie,
\textit{On-demand single-electron transfer between distant quantum dots},
Nature \textbf{477}, 439 (2011).

\bibitem{ford17}C. J. B. Ford,
\textit{Transporting and manipulating single electrons in surface-acoustic-wave minima},
Phys. Status Solidi B \textbf{254}, 1600658 (2017)

\bibitem{lima05}M. M. de Lima and P. V. Santos,
\textit{Modulation of photonic structures by surface acoustic waves},
Rep. Prog. Phys. \textbf{68}, 1639 (2005).

\bibitem{lima03}M. M. de Lima, F. Alsina, W. Seidel, and P. V. Santos,
\textit{Focusing of surface-acoustic-wave fields on (100) GaAs surfaces},
J. App. Phys. \textbf{94}, 7848 (2003).

\bibitem{ding13}X. Ding \textit{et al.},
\textit{Surface acoustic wave microfluidics},
Lab Chip \textbf{13}, 3626 (2013).

\bibitem{morgan07}D. Morgan, \textit{Surface Acoustic Wave Filters} (Academic
Press, Boston, 2007).

\bibitem{datta86}S. Datta, \textit{Surface Acoustic Wave Devices} (Prentice-Hall,
Upper Saddle River, NJ, 1986).

\bibitem{cerda-mendez10}E. A. Cerda-Mendez \textit{et al.},
\textit{Polariton condensation in dynamic acoustic lattices},
Phys. Rev. Lett. \textbf{105}, 116402 (2010).

\bibitem{kukushkin04}I. V. Kukushkin, J. H. Smet, L. Höppel, U. Waizmann,
M. Riek, W. Wegschneider, and K. von Klitzing,
\textit{Ultrahigh-frequency surface acoustic waves for finite wave-vector spectroscopy of two-dimensional electrons},
Appl. Phys. Lett. \textbf{85}, 4526 (2004).

\bibitem{byrnes07}T. Byrnes, P. Recher, N. Y. Kim, S. Utsunomiya,
and Y. Yamamoto,
\textit{Quantum Simulator for the Hubbard Model with Long-Range Coulomb Interactions Using Surface Acoustic Waves},
Phys. Rev. Lett. \textbf{99}, 016405 (2007).

\bibitem{buy12}S. Büyükköse,
B. Vratzov, D. Atac, J. van der Veen, P. V. Santos, and W.G. van der
Wiel,
\textit{Ultrahigh-frequency surface acoustic wave transducers on ZnO/$\textit{SiO}_2$/Si using nanoimprint lithography},
Nanotechnology \textbf{23}, 315303 (2012).

\bibitem{hutson62}A. R. Hutson and D. L. White,
\textit{Elastic Wave Propagation in Piezoelectric Semiconductors},
J. Appl. Phys. \textbf{33}, 40 (1962).

\bibitem{bierbaum72}P. Bierbaum,
\textit{Interaction of ultrasonic surface waves with conduction electrons in thin metal films},
Appl. Phys. Lett. \textbf{21}, 595 (1972).

\bibitem{wixforth86}A. Wixforth, J. P. Kotthaus, and G. Weimann,
\textit{Quantum Oscillations in the Surface-Acoustic-Wave Attenuation Caused by a Two-Dimensional Electron System},
Phys. Rev. Lett. \textbf{56}, 2104 (1986).

\bibitem{wixforth89}A. Wixforth, J. Scriba, M. Wassermeier, J. P.
Kotthaus, G. Weimann, and W. Schlapp,
\textit{Surface acoustic waves on GaAs/$\textit{Al}_x$$\textit{Ga}_{1-x}$As heterostructures},
Phys. Rev. B \textbf{40}, 7874 (1989).

\bibitem{paul90}W. Paul,
\textit{Electromagnetic traps for charged and neutral particles},
Rev. Mod. Phys. \textbf{62}, 531 (1990).

\bibitem{leibfried03}D. Leibfried, R. Blatt, C. Monroe, and D. Wineland,
\textit{Quantum dynamics of single trapped ions},
Phys. Rev. Mod. \textbf{75}, 281 (2003).

\bibitem{rahav03}S. Rahav, I. Gilary, and S. Fishman,
\textit{Time Independent Description of Rapidly Oscillating Potentials},
Phys. Rev. Lett. \textbf{91}, 110404 (2003).

\bibitem{rahav03b}S. Rahav, I. Gilary, and S. Fishman,
\textit{Effective Hamiltonians for periodically driven systems},
Phys. Rev. A \textbf{68}, 013820 (2003).

\bibitem{cirac94}J. I. Cirac, L. J. Garay, R. Blatt, A. S. Parkins,
and P. Zoller,
\textit{Laser cooling of trapped ions: The influence of micromotion},
Phys. Rev. A \textbf{49}, 421 (1994).

\bibitem{kohler97}S. Kohler, T. Dittrich, and P. Hänggi,
\textit{Floquet-Markovian description of the parametrically driven, dissipative harmonic quantum oscillator},
Phys. Rev. E \textbf{55}, 300 (1997).

\bibitem{fujisawa98}T. Fujisawa, T. H. Oosterkamp, W. G. van der
Wiel, B. W. Broer, R. Aguado, S. Tarucha, and L. P. Kouwenhoven,
\textit{Spontaneous Emission Spectrum in Double Quantum Dot Devices},
Science \textbf{282}, 932 (1998).

\bibitem{fujisawa02}T. Fujisawa, D. G. Austing, Y. Tokura, Y. Hirayama,
and S. Tarucha,
\textit{Allowed and forbidden transitions in artificial hydrogen and helium atoms},
Nature \textbf{419}, 278 (2002).

\bibitem{hayashi03}T. Hayashi, T. Fujisawa, H. D. Cheong, Y. H. Jeong,
and Y. Hirayama,
\textit{Coherent manipulation of electronic States in a double quantum dot},
Phys. Rev. Lett. \textbf{91}, 226804 (2003).

\bibitem{petta04}J. R. Petta, A. C. Johnson, C. M. Marcus, M. P.
Hanson, and A. C. Gossard,
\textit{Manipulation of a single charge in a double quantum dot},
Phys. Rev. Lett. \textbf{93}, 186802 (2004).

\bibitem{barthelemy13}P. Barthelemy and L. M. K. Vandersypen,
\textit{Quantum Dot Systems: a versatile platform for quantum simulations},
Ann. Phys. \textbf{525}, 808 (2013).

\bibitem{kornich14}V. Kornich, C. Kloeffel, and D. Loss,
\textit{Phonon-mediated decay of singlet-triplet qubits in double quantum dots},
Phys. Rev. B \textbf{89}, 085410 (2014).

\bibitem{wang13}K. Wang, C. Payette, Y. Dovzhenko, P. W. Deelman, and J. R. Petta,
\textit{Charge Relaxation in a Single Electron Si/SiGe Double Quantum Dot},
Phys. Rev. Lett. \textbf{111}, 046801 (2013).

\bibitem{rodriguez12}J. G. Rodriguez-Madrid, G. F. Iriarte, J. Pedros, O. A. Williams, D. Brink, and F. Calle,
\textit{Super-High-Frequency SAW Resonators on AlN/Diamond},
IEEE Electron Device Lett. \textbf{33}, 495 (2012).

\bibitem{benetti05}M. Benetti, D. Cannata, F. Di Pietrantonio, and E. Verona,
\textit{Growth of AlN Piezoelectric Film on Diamond for High-Frequency Surface Acoustic Wave Devices},
IEEE Trans. Ultrason. Ferroelectr. Freq. Control \textbf{52}, 1806 (2005).

\bibitem{assouar07}M. B. Assouar, O. Elmazria, P. Kirsch, P. Alnot, V. Mortet, and C. Tiusan,
\textit{High-frequency surface acoustic wave devices based on AlN/diamond
layered structure realized using e-beam lithography},
Journal of Applied Physics \textbf{101}, 114507 (2007).

\bibitem{glushkov12}E. Glushkov, N. Glushkova, and C. Zhang, 
\textit{Surface and pseudo-surface acoustic waves piezoelectrically excited in diamond-based structures},
Journal of Applied Physics \textbf{112}, 064911 (2012).

\bibitem{benetti05b}M. Benetti, D. Cannata, F. Di Pietrantonio, V. I. Fedosov, and E. Verona,
\textit{Gigahertz-range electro-acoustic devices based on pseudo-surface-acoustic waves in AlN/diamond/Si structures},
Appl. Phys. Lett. \textbf{87}, 033504 (2005). 

\bibitem{blick98}R. H. Blick, M. L. Roukes, W. Wegscheider, and M. Bichler, 
\textit{Freely suspended two-dimensional electron gases}, 
Physica B \textbf{249}, 784 (1998).

\bibitem{comsol}COMSOL Multiphysics\textregistered{} v. 5.2. www.comsol.com.
COMSOL AB, Stockholm, Sweden.

\bibitem{footnote-heterostructure}In our Comsol simulations we have neglected 
the presence of the thin $\mathrm{Al}_{x}\mathrm{Ga}_{1-x}\mathrm{As}$ crystal layer with typically $x\approx 0.3$. 
As argued in Ref.\cite{simon96}, this treatment is approximately correct
since the relevant material properties (elastic constants, densities, and dielectric constants) 
of $\mathrm{Al}_{x}\mathrm{Ga}_{1-x}\mathrm{As}$ and GaAs are very similar. 
The mode functions and speed of sound are largely defined by the elastic constants, 
which are roughly the same for both $\mathrm{Al}_{x}\mathrm{Ga}_{1-x}\mathrm{As}$ and pure GaAs; 
for example, the speed of the Rayleigh SAW mode for $\mathrm{Al}_{0.3}\mathrm{Ga}_{0.7}\mathrm{As}$ 
is $v_{s} \approx 3010\mathrm{m/s}$, which differs from that of pure GaAs by only $\sim 5\%$.
Also, the piezoelectric coupling constants are rather similar, with $e_{14} \approx 0.15 \mathrm{C/m^2}$ for pure 
GaAs and $e_{14} \approx 0.145 \mathrm{C/m^2}$ for $\mathrm{Al}_{0.3}\mathrm{Ga}_{0.7}\mathrm{As}$ \cite{simon96}. 

\bibitem{novoselov16}K. S. Novoselov \textit{et al.},
\textit{2D materials and van der Waals heterostructures},
Science \textbf{353}, aac9439 (2016).

\bibitem{kormanyos15}A. Kormanyos, G. Burkard, M. Gmitra, J. Fabian, V. Zolyomi, N. D. Drummond, and V. Fal'ko,
\textit{$k \cdot p$ theory for two-dimensional transition metal dichalcogenide semiconductors},
2D Mater. \textbf{2}, 022001 (2015)

\bibitem{eknapakul14}T. Eknapakul \textit{et al.},
\textit{Electronic Structure of a Quasi-Freestanding $\textit{MoS}_2$ Monolayer},
Nano Lett. \textbf{14}, 1312 (2014).

\bibitem{mak13}K. F. Mak, K. He, C. Lee, G. H. Lee, J. Hone, T. F. Heinz, and J. Shan,
\textit{Tightly bound trions in monolayer $\textit{MoS}_2$},
Nature Mat. \textbf{12}, 207 (2013). 

\bibitem{ross13}J. S. Ross \textit{et al.},
\textit{Electrical control of neutral and charged excitons in a monolayer semiconductor},
Nature Comm. \textbf{4}, 1474 (2013). 

\bibitem{srivastava15}A. Srivastava,	M. Sidler, A. V. Allain, D. S. Lembke,	A. Kis, and A. Imamoglu,
\textit{Optically active quantum dots in monolayer $\textit{WSe}_2$},
Nature Nano. \textbf{10}, 491 (2015). 

\bibitem{dawson1}P.H. Dawson, Y. Bingqi,
\textit{The second stability region of the quadrupole mass filter. I. Ion optical properties},
Int. J. Mass Spectrom. Ion Processes \textbf{\textcolor{black}{54}}, 25 (1983).

\bibitem{dawson2}P.H. Dawson, Y. Bingqi,
\textit{The second stability region of the quadrupole mass filter. I. Experimental results},
Int. J. Mass Spectrom. Ion Processes \textbf{\textcolor{black}{54}}, 41 (1983).

\bibitem{schuelein15}F. J. R. Schülein, E. Zallo, P. Atkinson, O.
G. Schmidt, R. Trotta, A. Rastelli, A. Wixforth, H. J. Krenner,
\textit{Fourier synthesis of radiofrequency nanomechanical pulses with different shapes},
Nature Nanotechn. \textbf{10}, 512 (2015).

\bibitem{possa16}G. C. Possa, L. F. Roncaratti,
\textit{Stability Diagrams for Paul Ion Traps Driven by Two Frequencies},
J. Phys. Chem. A \textbf{\textcolor{black}{120}}, 4915 (2016). 

\bibitem{footnote-amplitude}Here, we restrict our discussion 
to the pseudopotential regime where $q^{2}\ll1$, as done
in the major body of our work (apart from the discussion of exotic stability regions
where the heating-related arguments should be contrasted with an increased
temperature robustness).

\bibitem{furuta04}S. Furuta, C. H. W. Barnes, and C. J. L. Doran,
\textit{Single-qubit gates and measurements in the surface acoustic wave quantum computer},
Phys. Rev. B \textbf{70}, 205320 (2004).

\bibitem{martins16}F. Martins, F. K. Malinowski, P. D. Nissen, E.
Barnes, S. Fallahi, G. C. Gardner, M. J. Manfra, C. M. Marcus, and
F. Kuemmeth, 
\textit{Noise Suppression Using Symmetric Exchange Gates in Spin Qubits},
Phys. Rev. Lett. \textbf{116}, 116801 (2016). 

\bibitem{reed16}M. D. Reed \textit{et al.}, 
\textit{Reduced Sensitivity to Charge Noise in Semiconductor Spin Qubits via Symmetric Operation},
Phys. Rev. Lett. \textbf{116}, 110402 (2016). 

\bibitem{baart17}T. A. Baart, T. Fujita, C. Reichl, W. Wegscheider,
and L. M. K. Vandersypen, 
\textit{Coherent spin-exchange via a quantum mediator},
Nature Nano. \textbf{12}, 26 (2017). 

\bibitem{campbell89}C. Campbell, \textit{Surface Acoustic Wave Devices
for Mobile and Wireless Communications} (Academic Press, 1998). 

\bibitem{jaksch98}D. Jaksch \textit{et al.},
\textit{Cold bosonic atoms in optical lattices},
Phys. Rev. Lett. \textbf{81}, 3108 (1998).

\bibitem{hofstetter02}W. Hofstetter, J. I. Cirac, P. Zoller, E. Demler,
and M. D. Lukin,
\textit{High-temperature superfluidity of fermionic atoms in optical lattices},
Phys. Rev. Lett. \textbf{89}, 220407 (2002).

\bibitem{anderson75}P. W. Anderson,
\textit{Model for the Electronic Structure of Amorphous Semiconductors},
Phys. Rev. Lett. \textbf{34}, 953 (1975).

\bibitem{singha11}A. Singha \textit{et al.},
\textit{Two-Dimensional Mott-Hubbard Electrons in an Artificial Honeycomb Lattice},
Science \textbf{332}, 1176 (2011).

\bibitem{schlosser96}T. Schl\"osser \textit{et al.},
\textit{Internal structure of a Landau band induced by a lateral superlattice: a glimpse of Hofstadter's butterfly},
Eur. Phys. Lett. \textbf{33}, 683 (1996).

\bibitem{albrecht01}C. Albrecht \textit{et al.},
\textit{Evidence of Hofstadters Fractal Energy Spectrum in the Quantized Hall Conductance},
Phys. Rev. Lett. \textbf{86}, 147 (2001).

\bibitem{hensgens17}T. Hensgens, T. Fujita, L. Janssen, Xiao Li,
C. J. Van Diepen, C. Reichl, W. Wegscheider, S. Das Sarma, and L.
M. K. Vandersypen,
\textit{Quantum simulation of a Fermi-Hubbard model using a semiconductor quantum dot array},
Nature \textbf{548}, 73 (2017).

\bibitem{kuljanishvili08}I. Kuljanishvili \textit{et al.},
\textit{Scanning-probe spectroscopy of semiconductor donor molecules},
Nature Phys. \textbf{4}, 227 (2008).

\bibitem{martin04}J. Martin, S. Ilani, B. Verdene, J. Smet, V. Umansky, D. Mahalu, D. Schuh, G. Abstreiter, and A. Yacoby, 
\textit{Localization of Fractionally Charged Quasi-Particles},
Science \textbf{305}, 980 (2004).

\bibitem{dial07}O. E. Dial, R. C. Ashoori, L. N. Pfeiffer, and K. W. West,
\textit{High-resolution spectroscopy of two-dimensional electron systems},
Nature \textbf{448}, 176 (2007). 

\bibitem{gywat10}O. Gywat, H. J. Krenner, and J. Berezovsky,
\textit{Spins in optically active quantum dots}, Wiley-VCH (2010).

\bibitem{vamivakas09}A. N. Vamivakas, Y. Zhao, C.-Y. Lu, and M. Atat\"ure,
\textit{Spin-resolved quantum-dot resonance fluorescence},
Nature Phys. \textbf{5}, 198 (2009).

\bibitem{atatuere07}M. Atat\"ure, J. Dreiser, A. Badolato, and A. Imamoglu,
\textit{Observation of Faraday rotation from a single confined spin},
Nature Phys. \textbf{3}, 101 (2007).

\bibitem{berezovsky06}J. Berezovsky, M. H. Mikkelsen, O. Gywat, N. G. Stoltz, L. A. Coldren, and D. D. Awschalom,
\textit{Nondestructive Optical Measurements of a Single Electron Spin in a Quantum Dot},
Science \textbf{314}, 1916 (2006).

\bibitem{baart16}T. A. Baart, M. Shafiei, T. Fujita, C. Reichl, W. Wegscheider, and L. M. K. Vandersypen,
\textit{Single-spin CCD},
Nature Nano. \textbf{11}, 330 (2016).

\bibitem{byrnes08}T. Byrnes, N. Kim, K. Kusudo, and Y. Yamamoto,
\textit{Quantum simulation of Fermi-Hubbard models in semiconductor quantum-dot arrays},
Phys. Rev. B \textbf{78}, 075320 (2008).

\bibitem{footnote-mobility}Here, we take the mobility $\mu$ as our
figure of merit,  as it is one of the standard metrics to characterize the
effect of disorder. Both long-range scattering and short-range scattering
are reflected in the mobility and in fact the exact dependence of
mobility on density often allows one to distinguish whether long-range
or short-range scattering dominates. The main caveat about using mobility
is that it is typically measured in density ranges larger than those
used in the present experiments. The same caveat, however, applies to the use
of mobility for predicting quantum dot behaviour. Yet, empirically,
for quantum dots defined in systems with relatively low mobility,
such as Si/SiGe quantum wells, mobility is found to be a good predictor
of the ability to realize well-behaved quantum dots, i.e. not suffering
excessive disorder \cite{zwanenburg13}.
For this reason, and given that mobility has been measured across
many systems, we refer to mobility as an indication of disorder.

\bibitem{umansky97}V. Umansky, R. de-Picciotto, and M. Heiblum,
\textit{Extremely high-mobility two dimensional electron gas: Evaluation of scattering mechanisms},
Appl. Phys. Lett. \textbf{71}, 683 (1997).

\bibitem{hofstadter76}D. Hofstadter,
\textit{Energy levels and wave functions of Bloch electrons in rational and irrational magnetic fields},
Phys. Rev. B \textbf{14}, 2239 (1976).

\bibitem{lagendijk09}A. Lagendijk, B. A. Van Tiggelen, and D. Wiersma,
\textit{Fifty years of Anderson localization},
Phys. Today \textbf{62}, 24 (2009).

\bibitem{belitz94}D. Belitz and T. R. Kirkpatrick,
\textit{The Anderson-Mott transition},
Rev. Mod. Phys. \textbf{66}, 261 (1994).

\bibitem{basko06}D. Basko, I. Aleiner, and B. Altshuler,
\textit{Metal-insulator transition in a weakly interacting many-electron system with localized single-particle states},
Ann. Phys. \textbf{321}, 1126 (2006).

\bibitem{byczuk05}K. Byczuk, W. Hofstetter, and D. Vollhardt,
\textit{Mott-Hubbard Transition versus Anderson Localization in Correlated Electron Systems with Disorder},
Phys. Rev. Lett. \textbf{94}, 056404 (2005).

\bibitem{fallani07}L. Fallani, J. E. Lye, V. Guarrera, C. Fort, and
M. Inguscio,
\textit{Ultracold Atoms in a Disordered Crystal of Light: Towards a Bose Glass},
Phys. Rev. Lett. \textbf{98}, 130404 (2007).

\bibitem{sushkov13}O. P. Sushkov and A.H. Castro Neto,
\textit{Topological Insulating States in Laterally Patterned Ordinary Semiconductors},
Phys. Rev. Lett. \textbf{110}, 186601 (2013).

\bibitem{buller16}J. V. T. Buller, R. E. Balderas-Navarro, K. Biermann, E. A. Cerda-Mendez, and P. V. Santos, 
\textit{Exciton-polariton gap soliton dynamics in moving acoustic square lattices},
Phys. Rev. B \textbf{94}, 125432 (2016). 

\bibitem{butov04}L. V. Butov \textit{et al.},
\textit{Condensation and pattern formation in cold exciton gases in coupled quantum wells},
J. Phys. Condens. Matter \textbf{16}, R1577 (2004).

\bibitem{hammack06}A. T. Hammack \textit{et al.},
\textit{Trapping of Cold Excitons with Laser Light},
Phys. Rev. Lett. \textbf{96}, 227402 (2006).

\bibitem{hammack07}A. T. Hammack \textit{et al.},
\textit{Kinetics of indirect excitons in an optically induced trap in GaAs quantum wells},
Phys. Rev. B \textbf{76}, 193308 (2007).

\bibitem{kumar15}S. Kumar, A. Kaczmarczyk, and B. D. Gerardot,
\textit{Strain-Induced Spatial and Spectral Isolation of Quantum Emitters in Mono- and Bilayer $\textit{WSe}_2$},
Nano Lett. \textbf{15}, 7567 (2015).

\bibitem{schuetz15}M. J. A. Schuetz, E. M. Kessler, G. Giedke, L.
M. K. Vandersypen, M. D. Lukin, and J. I. Cirac,
\textit{Universal Quantum Transducers Based on Surface Acoustic Waves},
Phys. Rev. X \textbf{5}, 031031 (2015).

\bibitem{liu08}J.-X. Liu, D.-N. Fang, W.-Y. Wei, and X.-F. Zhao,
\textit{Love waves in layered piezoelectric/piezomagnetic structures},
J. Sound Vib. \textbf{315}, 146 (2008).

\bibitem{pang08}Y. Pang, J.-X. Liu, Y.-S. Wang, and X.-F. Zhang,
\textit{Propagation of Rayleigh-type surface waves in a transversely isotropic piezoelectric layer on a piezomagnetic half-space},
J. Appl. Phys. \textbf{103}, 074901 (2008).

\bibitem{weiler11}M. Weiler, L. Dreher, C. Heeg, H. Huebl, R. Gross,
M. S. Brandt, and S. T. B. Goennenwein,
\textit{Acoustically driven ferromagnetic resonance},
Phys. Rev. Lett. \textbf{106}, 117601 (2011).

\bibitem{knoerzer17}J. Kn\"orzer \textit{et al.}, in preparation.

\bibitem{glauber92}R. J. Glauber, in Laser Manipulation of Atoms
and Ions, Proceedings of the International School of Physics ``Enrico
Fermi'' Course 118, edited by E. Arimondo, W. D. Phillips, and F.
Strumia (North-Holland, Amsterdam), p. 643 (1992).

\bibitem{golovach04}V. N. Golovach, A. Khaetskii, and D. Loss,
\textit{Phonon-Induced Decay of the Electron Spin in Quantum Dots},
Phys. Rev. Lett. \textbf{93}, 016601 (2004).

\bibitem{breuer02}H.-P. Breuer and F. Petruccione, \textit{The Theory
of Open Quantum Systems} (Oxford University Press, Oxford, 2002).

\bibitem{schnebele06}R. J. Schnebele \textit{et al.}, 
\textit{Quantum-dot thermometry of electron heating by surface acoustic waves},
Appl. Phys. Lett. \textbf{89}, 122104 (2006). 

\bibitem{utko06}P. Utko, P. E. Lindelof, and K. Gloos, 
\textit{Heating in single-electron pumps driven by surface acoustic waves},
Appl. Phys. Lett. \textbf{88}, 202113 (2006).

\bibitem{manenti16}R. Manenti, M. J. Peterer, A. Nersisyan, E. B.
Magnusson, A. Patterson, and P. J. Leek, 
\textit{Surface acoustic wave resonators in the quantum regime},
Phys. Rev. B \textbf{93}, 041411(R) (2016).

\bibitem{wu14}W. Wu \textit{et al.}, 
\textit{Piezoelectricity of single-atomic-layer MoS2 for energy conversion and piezotronics},
Nature \textbf{514}, 470 (2014).

\bibitem{ekstroem17}M. K. Ekstroem, T. Aref, J. Runeson, J. Bjoerck, I. Bostroem, and P. Delsing, 
\textit{Surface acoustic wave unidirectional transducers for quantum applications},
Appl. Phys. Lett. \textbf{110}, 073105 (2017).

\bibitem{vandersypen16}L. M. K. Vandersypen, H. Bluhm, J. S. Clarke,
A. S. Dzurak, R. Ishihara, A. Morello, D. J. Reilly, L. R. Schreiber, and M. Veldhorst, 
\textit{Interfacing spin qubits in quantum dots and donors - hot, dense and coherent},
arXiv:1612.05936.

\bibitem{chekhovich13}E. A. Chekhovich, M. N. Makhonin, A. I. Tartakovskii,
A. Yacoby, H. Bluhm, K. C. Nowack, and L. M. K. Vandersypen,
\textit{Nuclear spin effects in semiconductor quantum dots},
Nat. Mater. \textbf{12}, 494 (2013).

\bibitem{zwanenburg13}F. A. Zwanenburg, A. S. Dzurak, A. Morello,
M. Y. Simmons, L. C. L. Hollenberg, G. Klimeck, S. Rogge, S. N. Coppersmith,
and M. A. Eriksson, Silicon Quantum Electronics,
\textit{Silicon quantum electronics},
Rev. Mod. Phys. \textbf{85}, 961 (2013).

\bibitem{veldhorst14}M. Veldhorst\textit{ et al.},
\textit{An addressable quantum dot qubit with fault-tolerant control fidelity},
Nat. Nanotechnol. \textbf{9}, 981 (2014).

\bibitem{simon96}S. H. Simon,
\textit{Coupling of surface acoustic waves to a two-dimensional electron gas},
Phys. Rev. B \textbf{54}, 13878 (1996).

\end{thebibliography}
\end{document}